\documentclass[twocolumn]{aastex701}

\usepackage{amsmath,amssymb}
\usepackage{graphicx}
\usepackage{booktabs}
\graphicspath{{figures/}}

\newcommand{\jaisp}{\textsc{Jaisp}}
\newcommand{\arcsecpx}{\arcsec\,\mathrm{px}^{-1}}

\newcommand{\ipac}{\affiliation{IPAC, California Institute of Technology, Pasadena, CA 91125, USA}}
\newcommand{\JPL}{\affiliation{Jet Propulsion Laboratory, California Institute of Technology, 4800 Oak Grove Drive, Pasadena, CA 91109, USA}}

\usepackage{xcolor}

\shorttitle{JAISP: A Rubin–Euclid Foundation Model}
\shortauthors{Hemmati et al.}

\begin{document}

\title{JAISP: A Joint Rubin–Euclid Foundation Model for Cross-Instrument Source Detection and Astrometry}

\author[0000-0003-2226-5395]{Shoubaneh Hemmati}
\ipac
\email{shemmati@caltech.edu}

\author[0000-0003-3367-3415]{George Helou}
\ipac
\email{}

\author[0000-0001-8792-3091]{Yu-Heng Lin}
\ipac
\email{}

\author[0000-0001-9674-1564]{Rachel Akeson}
\ipac
\email{}

\author[0009-0002-0098-6183]{Charles Darren Dowell}
\JPL
\email{}

\author[0000-0002-9382-9832]{Andreas Faisst}
\ipac
\email{}

\author[0000-0002-8963-8056]{Thomas Greene}
\ipac
\email{}

\author[0000-0003-4401-0430]{David Shupe}
\ipac
\email{}

\begin{abstract}
Joint pixel-level processing of Rubin and \textit{Euclid} imaging promises deeper source detection and the cross-survey consistency that precision cosmology requires. We introduce \jaisp{} (Joint AI Survey Processing), a foundation model that learns a single representation of ten-band imaging: Rubin $ugrizy$ and \textit{Euclid} VIS, $Y$, $J$, and $H$, from the first public releases over ${\approx}0.2$\,deg$^2$ of the ECDFS deep field. Trained without labels by predicting each band from the other nine, its frozen encoder drives small task-specific networks for source detection and astrometry. The detection network recovers nearly all VIS-selected \textit{Euclid} Q1 catalog sources on held-out sky, and source injection shows that combining all ten bands extends the point-source detection limit by ${\approx}0.3$\,mag beyond VIS alone. The astrometry network transfers the sharp VIS localization to the other nine bands, reducing median cross-band positional offsets from ${\approx}50$ to $6$--$10$\,mas while retaining the full precision of its VIS centroid anchor even for faint sources. Separating this per-source scatter exposes the cross-survey concordance field: a coherent $9$--$10$\,mas offset between the two independently \textit{Gaia}-registered astrometric solutions, dominated by a component common to all bands. Applied without retraining to a second \textit{Euclid} Deep Field, both networks perform without measurable loss. These gains serve the joint-survey science case: a deeper census recovers faint neighbors whose blending biases weak-lensing shapes and photometric redshifts, and VIS-grade positions in every band supply the common astrometric reference that joint photometry and faint-source proper motions require. A single representation learned once from survey pixels thus serves several measurements across the sky.
\end{abstract}


\section{Introduction}

Cosmology has entered a regime where its hardest questions are limited by systematics rather than by statistics. The current generation of wide-field surveys is measuring the growth of structure and the expansion history precisely enough that the residual tensions between probes can no longer be dismissed as noise, yet cannot be confidently attributed to new physics rather than to unmodeled measurement error \citep{Rhodes2017}. Breaking this impasse depends on the next generation of surveys not merely collecting more photons, but controlling systematics far more tightly than before.

The Vera C. Rubin Observatory's Legacy Survey of Space and Time \citep[LSST;][]{Ivezic2019}, ESA's \textit{Euclid} mission \citep{Mellier2025}, and the \textit{Nancy Grace Roman Space Telescope} \citep{Spergel2015} will observe overlapping sky from ${\sim}0.3$ to 2~\micron\ at sub-arcsecond resolution. Individually powerful, the observatories in combination can calibrate and cross-check one another, suppressing survey-specific systematics that no single survey can control alone. The case for combining them at the pixel level rather than the catalog level was laid out by the Joint Survey Processing study \citep{Chary2020}. At these depths up to a quarter of the sources in first-year ground-based stacks, and up to half at full depth, are blended at seeing-limited resolution, and astrometric scatter between surveys propagates directly into the error of any joint photometry. Both effects feed the weak-lensing shape and photometric-redshift measurements at the core of the cosmology case, and the same cross-survey positions, for sources far fainter than \textit{Gaia} reaches, enable proper motions and astrometric microlensing for Galactic science. That study therefore placed a common astrometric and photometric reference across the surveys, its concordance frame, first among its requirements. In addition to cosmology, joint processing benefits many other astrophysical pursuits, including reionization and galaxy evolution \citep{Faisst2022}, microlensing by free-floating planets and Galactic black holes, stellar proper motions \citep{FajardoAcosta2022}, and solar system objects \citep{Chary2020}. Exploiting the overlap at the pixel level, however, means confronting heterogeneous PSFs, samplings, and noise properties simultaneously, a regime where classical pipelines strain.

Machine and deep learning have moved quickly from peripheral tools to central ones in survey astronomy \citep[e.g.,][]{baron2019, Huertas2023, Smith_2023}, and there is a growing case that much of the next decade of pixel-level processing will rely on learned components alongside explicitly modeled ones. This is not, however, a shortcut. The classical pipelines these surveys rely on represent decades of accumulated, problem-specific effort: careful PSF models, distortion solutions, deblenders, and uncertainty propagation, each refined against the failure modes of real data. Reaching that level of fidelity with learned methods requires comparable investment, because the understanding encoded in the classical solution is precisely what tells us how to design, constrain, and validate its learned replacement, and sets the benchmark to meet and surpass. Once that effort is invested, the payoff is a more data-driven pipeline: inference is fast, and the representation is shaped by the pixels themselves rather than by parametric assumptions, absorbing structure such as cross-instrument PSF differences, wavelength-dependent morphology, and correlated noise that classical stages can only approximate.

This does not mean every problem calls for a single, large foundation model. Much recent work comes instead from compact, purpose-built networks designed around a specific measurement and its systematics: generative and physics-informed models for deblending, super-resolution, and the realistic image simulations needed for shear calibration \citep[e.g.,][]{Hemmati2022, Haghjoo2026, Scognamiglio2025, Scognamiglio2026}. In these, the architecture follows directly from the problem. Foundation models occupy a complementary niche: rather than solving one task, they learn a general representation once through self-supervised pretraining and amortize it across many downstream tasks. This paradigm has begun to take hold in astronomy, through cross-modal and autoregressive models trained on large galaxy samples and the multi-survey datasets assembled to support them \citep[e.g.,][]{Parker_2024, Smith2024, MMU2024}. Building on this direction, we apply it to a problem central to joint survey processing: combining three imaging instruments at the pixel level, across mismatched resolutions and overlapping but non-identical bandpasses.

In this work we introduce \jaisp{} (Joint AI Survey Processing), which learns a single shared representation of Rubin and \textit{Euclid} imaging, using only the first public data release of each, through self-supervised pretraining, and reuses it across several downstream measurements. We work from already-processed survey mosaics rather than individual exposures, which lets us concentrate on the joint representation itself; the same approach could in principle be pushed earlier in the pipeline, toward less-processed data where more of the instrumental signal survives. A foundation model is trained once to capture the cross-instrument spatial and spectral structure that joint processing requires; its frozen features then drive lightweight task-specific heads. We focus here on source detection and astrometry, benchmarked against classical baselines; the same frozen features can serve many further downstream measurements, including photometry, shape measurement, and redshift estimation, which we develop in forthcoming work. Even at the mosaic level, we find that the learned representation transfers \textit{Euclid}'s sharp localization to Rubin's bands and to faint sources, improves source detection by combining all ten bands, and measures the relative astrometric offsets between two independently \textit{Gaia}-aligned surveys. These results hold on a separate deep field with no retraining, a first test of the representation's transferability across the sky.

This paper is organized as follows. Section~\ref{sec:data} describes the Rubin and \textit{Euclid} imaging; Section~\ref{sec:architecture} presents the foundation model and the frozen-feature head design, and validates the learned representation; Sections~\ref{sec:detection} and~\ref{sec:astrometry} present the detection and astrometry heads; Section~\ref{sec:ood} tests transfer to a separate deep field; Section~\ref{sec:release} describes the public code and data products; and Section~\ref{sec:discussion} summarizes the results and discusses implications and next steps.

\section{Data}\label{sec:data}
\begin{figure*}[t]
    \centering
    \includegraphics[width=\textwidth]{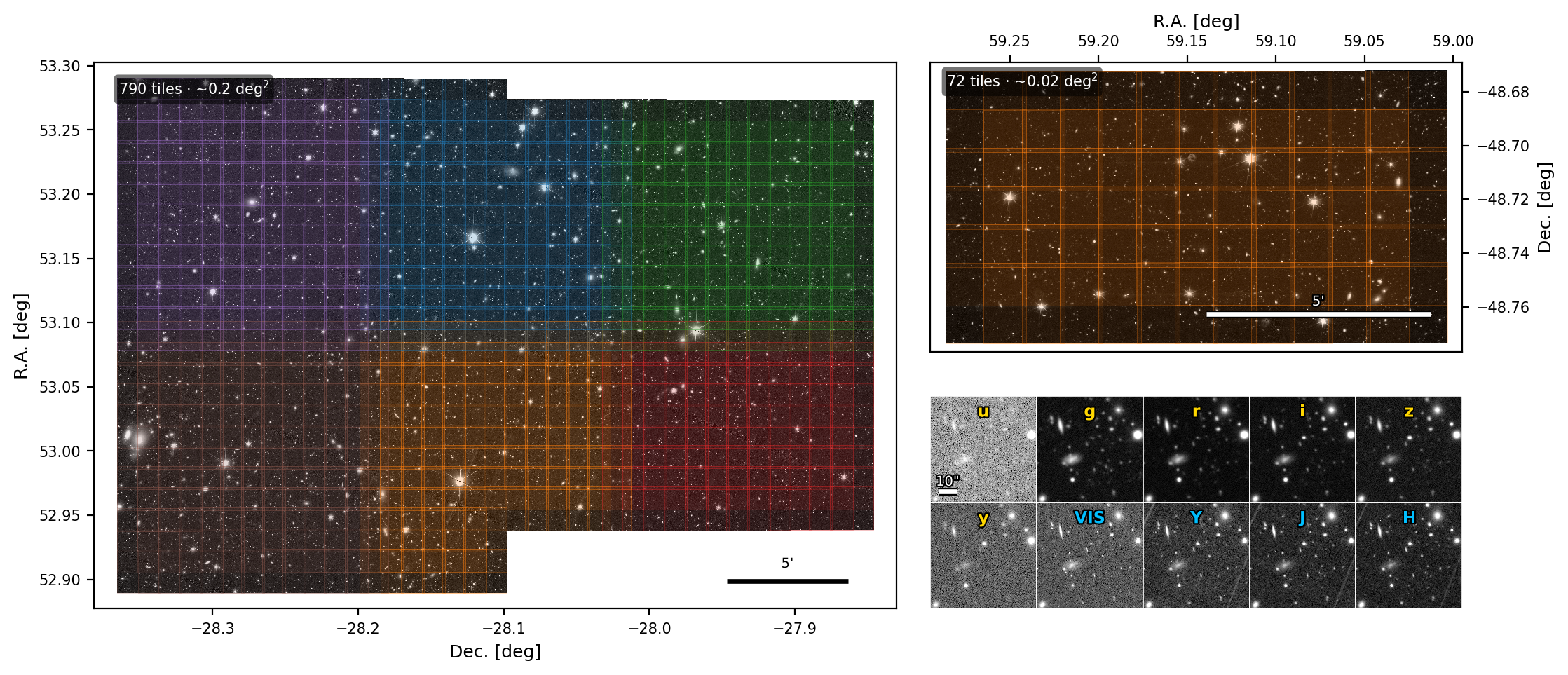}
\caption{Data and survey footprints. \emph{Left:} the ECDFS field used for training, where Rubin and \textit{Euclid} coverage overlap: a single Rubin tract (5063) spanning six patches, tiled into 790 $512\times512$ cutouts (stride 256\,px, 50\% overlap) over ${\sim}0.2$\,deg$^2$, shown on the \textit{Euclid} VIS mosaic and tinted by patch. \emph{Top right:} the EDF-S field, a widely separated region held out to test out-of-distribution transfer with no retraining: 72 tiles (${\sim}0.02$\,deg$^2$). \emph{Bottom right:} the ten input bands for one matched tile, Rubin $ugrizy$ (gold labels) and \textit{Euclid} VIS and NISP $Y$/$J$/$H$ (blue labels); each panel is a $1'$ field. Rubin imaging is sampled at $0.2\arcsecpx$ and the \textit{Euclid} bands are drawn from the $0.1\arcsecpx$ MER mosaics, giving the joint ten-band stack used throughout this work.}
    \label{fig:data}
\end{figure*}

The dataset for this work combines overlapping imaging from Rubin and \textit{Euclid} over two widely separated Euclid Deep Fields. Training uses the Euclid Deep Field Fornax (EDF-F), which coincides with the Extended Chandra Deep Field South (ECDFS) in the Rubin footprint; the Euclid Deep Field South (EDF-S) is reserved entirely for the out-of-distribution transfer test of Section~\ref{sec:ood} and is never seen during training. For each field we assemble a common ten-band stack: the six Rubin optical bands $ugrizy$ together with \textit{Euclid} VIS and the three NISP near-infrared bands $Y$, $J$, and $H$ (Figure~\ref{fig:data}). All imaging is drawn from currently public data releases. The region used here is a small part of the available Rubin--\textit{Euclid} overlap: it is a single Rubin tract within one deep field, while the deep overlap spans several tracts per field and multiple fields, leaving substantial room to scale the same approach to larger areas as deeper releases arrive.

Rubin imaging is drawn from the Legacy Survey of Space and Time Data Preview 1 \citep[DP1;][]{DP1}, the first release from the Rubin Observatory. DP1 comprises coadded $ugrizy$ imaging obtained with the Rubin Commissioning Camera (LSSTComCam) during on-sky commissioning in late 2024, covering seven non-contiguous fields. Of these fields, ECDFS is the deepest and most uniformly observed, and we use its coadds in all six bands. Within ECDFS we use a single Rubin tract (tract 5063), spanning six patches, which covers the region of common Rubin--\textit{Euclid} depth used here. The coadds are sampled at $0.2\arcsecpx$; the median point-spread function is broad by space-based standards, with a full width at half maximum of roughly $1''$ at this early commissioning stage. \textit{Euclid} VIS provides the sharp spatial reference throughout this work. The $5\sigma$ point-source coadd depths in this field reach ${\approx}26$ in $g$ and $r$, with $u\approx24.6$ and $y\approx23.1$ the shallowest; the full per-band values and processing details are given in the DP1 release paper.

\textit{Euclid} imaging is drawn from the Quick Data Release 1 \citep[Q1;][]{EuclidQ1}, the first public \textit{Euclid} release, which covers the three Euclid Deep Fields, including EDF-F and EDF-S used here. Despite the deep-field designation, Q1 provides a single visit of each field, reaching approximately the depth of the Euclid Wide Survey; the full deep-field depth will accumulate over repeated visits during the mission. \textit{Euclid} carries two instruments. The visible imager VIS \citep{Cropper2025} observes in a single broad band spanning ${\approx}550$--$900$\,nm at a native $0.1\arcsecpx$ sampling, and the Near-Infrared Spectrometer and Photometer \citep[NISP;][]{Jahnke2025} images in three bands, $Y$, $J$, and $H$, covering ${\approx}950$--$2020$\,nm \citep{Schirmer2022} at a native $0.3\arcsecpx$. We use the Q1 VIS \citep{VIS} and NISP \citep{NIR} imaging as delivered by the MERge processing function \citep[MER;][]{MER}, the pipeline stage that combines the calibrated frames of both instruments into background-subtracted mosaics on a common $0.1\arcsecpx$ grid \citep{IRSA601} and produces the multiwavelength source catalog \citep{IRSA602} used as a reference in Section~\ref{sec:detection}. On that grid VIS provides the highest resolution ($\sim0.16''$ FWHM) and the NISP bands are resampled from their coarser native sampling; the $5\sigma$ point-source depths are $\approx26$ in VIS and $\approx24$ in each NISP band. Per-pixel variance maps accompany every band and are used for noise normalization throughout (Section~\ref{sec:architecture}).

Using the MER mosaics means inheriting that preprocessing: background subtraction, and in particular the resampling of the native near-infrared NISP frames onto the finer VIS pixel grid, are applied before our pipeline sees the data. This is a deliberate, conservative starting point that keeps the input fully reproducible from public products. Q1 does already include the near-infrared mosaics at their native NISP sampling, but as science images alone, without the per-pixel variance maps our pipeline requires for noise normalization. Native-sampling products with their full supporting maps, expected in deeper releases beginning with DR1, would let a joint model ingest them directly and perform the cross-instrument registration itself, reducing the preprocessing carried over from the survey pipeline; we return to this in Section~\ref{sec:discussion}.

For each field we lay the imaging out on a common tile grid. Rubin tiles are $512\times512$ pixels ($\sim102''$ on a side at $0.2\arcsecpx$), with the matching \textit{Euclid} region cut at twice the sampling; the ten bands of each tile are spatially aligned to a common frame. Tiles are placed on a $256$-pixel stride, so adjacent tiles overlap by $50\%$ and a given point on the sky falls in up to four tiles. The training field yields $790$ matched Rubin--\textit{Euclid} tiles over $\sim0.2$\,deg$^2$; the held-out EDF-S field provides a further $72$ matched tiles ($\sim0.02$\,deg$^2$), used only for the transfer test of Section~\ref{sec:ood}. This overlap is useful during pretraining, where the same source appears at different positions relative to the tile edge and so provides free augmentation, and during detection, where sources that sit near the edge of one tile fall near the center of a neighbor.

Because neighboring tiles overlap, excluding a tile from training does not necessarily exclude all of the sky it contains: some pixels may also appear in adjacent training tiles. The evaluations in the following sections therefore use different levels of hold-out. Individual tiles are reserved for reconstruction diagnostics, an entire ECDFS patch is withheld when testing task-specific models, and the separate EDF-S field is excluded from all training to test transfer to new sky. These tests address different questions, so we specify which data are withheld at each training stage when presenting the corresponding results. Table~\ref{tab:evaluation} summarizes the samples and their relationship to training.

\section{Architecture}\label{sec:architecture}

\begin{figure*}[t]
    \centering
    \includegraphics[width=\textwidth]{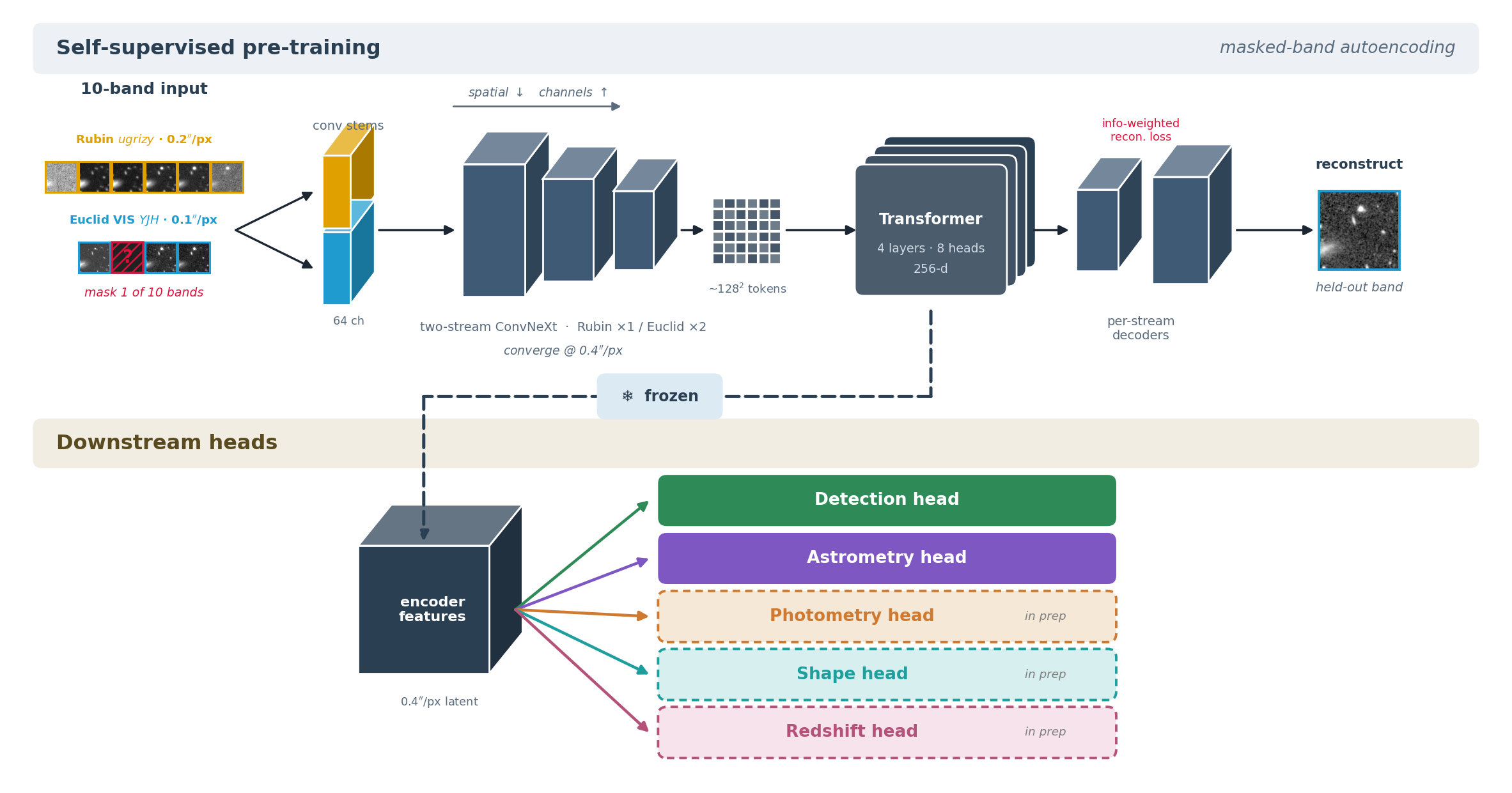}
\caption{Model architecture. \emph{Top:} self-supervised pretraining by masked-band autoencoding. The ten input bands, Rubin $ugrizy$ at $0.2\arcsecpx$ and \textit{Euclid} VIS\,$YJH$ at $0.1\arcsecpx$, pass through per-band convolutional stems and a two-stream ConvNeXt encoder (Rubin and \textit{Euclid} branches) whose feature maps shrink spatially and deepen in channels until they converge on a common $0.4\arcsecpx$ grid. A transformer bottleneck (4 layers, 8 heads, 256-d) mixes the ${\sim}128^2$ fused tokens, each carrying joint Rubin--\textit{Euclid} features for one $0.4\arcsecpx$ location, and per-stream decoders reconstruct the single held-out band under an information-weighted loss; the complete model totals ${\sim}9.1$M parameters. \emph{Bottom:} for downstream tasks the encoder is frozen and its $0.4\arcsecpx$ latent features feed lightweight task heads, detection and astrometry in this work, with photometry, shape, and redshift heads in preparation. Input and reconstruction thumbnails show the same matched ECDFS tile as Figure~\ref{fig:data}.}
    \label{fig:arch}
\end{figure*}

The core of \jaisp{} is a single foundation model that learns a shared representation of the Rubin and \textit{Euclid} imaging, which is then reused by separate task-specific heads. The foundation model is trained by self-supervised masked-band reconstruction, a masked autoencoder \citep[MAE;][]{MAE} applied across wavelength bands rather than spatial patches: from the multi-band input, one band is withheld and the model is required to predict its image from the remaining ones. No labels are used at this stage, since the supervision comes entirely from the requirement that the model reproduce held-out pixels, so the model can be trained on raw imaging alone. Once trained, the encoder is frozen and its features drive the downstream heads. This work develops the detection and astrometry heads, with photometry, shape-measurement, and redshift heads in preparation (Figure~\ref{fig:arch}).

The architecture must reconcile a mismatch between the instruments. Rubin imaging is sampled at $0.2\arcsecpx$ across six optical bands ($ugrizy$), whereas \textit{Euclid} VIS and the three NISP bands ($Y$, $J$, $H$) from MER are sampled twice as finely, at $0.1\arcsecpx$. Resampling both onto the coarser Rubin grid would reduce the sampling of the sharper VIS image. The model instead keeps each instrument at its delivered pixel sampling through separate encoder branches, and brings them together only once both have been reduced to the same angular sampling on the sky.

Each band first passes through its own convolutional ``stem,'' a small network that maps a single noisy band into a shared feature space while normalizing for that band's noise level. The per-band features then enter one of two parallel encoder branches, one for Rubin and one for \textit{Euclid}, built from ConvNeXt blocks \citep{liu2022,convnext}. ConvNeXt modernizes the classical convolutional network with design choices adapted from vision transformers \citep{ViT}, such as large depthwise kernels and inverted bottlenecks, matching transformer performance on vision benchmarks while remaining purely convolutional: it keeps the locality and translation equivariance natural to imaging data, and its cost scales linearly with image area rather than quadratically as self-attention does, which matters at the pixel counts of survey imaging. We therefore use convolutions wherever the sampling is fine and reserve explicit attention for the coarse fused grid below. Within each branch the feature maps are progressively reduced in spatial size and increased in depth, so that despite their different delivered sampling both branches arrive at a common $0.4\arcsecpx$ grid.

The two streams are then summed with learned identity embeddings that let the network tell Rubin features from \textit{Euclid} ones, and passed to a transformer bottleneck \citep[][]{transformers}. Through self-attention, every location in the fused map can exchange information with every other, and it is here that cross-instrument and cross-band relationships are combined. Per-instrument decoders expand the bottleneck features back to the delivered sampling to reconstruct the withheld band, with the target band specified to the decoder so that a single model can reconstruct any of the bands. The complete model contains roughly nine million parameters; layer-by-layer specifications are given in Figure~\ref{fig:arch}.

The $0.4\arcsecpx$ fused grid balances spatial sampling, sky coverage, and computational cost. We tested a finer $0.2\arcsecpx$ grid, halving the Rubin training-crop width from 256 to 128 pixels to keep the transformer token count fixed and training computationally tractable. At evaluation, both models were applied to identical sky crops at both sizes, allowing the finer grid four times as many tokens. Reconstruction was nearly unchanged at both crop sizes, and feature-decoding performance was comparable on the 256-pixel crops. These results support retaining the coarser grid for this work, although they do not test training the finer grid with the larger sky context.

Within each branch the per-band feature maps must be combined before the streams are fused. A natural default is to pool them into a single map, but doing so attenuates each band's contribution to the learning signal: averaging six Rubin bands, for example, divides the gradient passed to each input feature map by six. The model instead concatenates the per-band features and projects them, giving each band its own learned weights. This keeps wavelength-dependent information, such as how a galaxy's morphology and effective PSF change from band to band, available to the representation, where the photometry and deblending heads later draw on it. When a band is withheld as the reconstruction target its input slot is zero-filled, and the projection learns to ignore the empty slot, which also lets the model operate when some bands are absent.

The reconstruction loss is computed in noise-normalized units and weighted toward pixels carrying source signal, so the model is judged on how well it reproduces astronomical structure rather than blank sky. Without this weighting the loss would be dominated by the empty background that fills most of every image, and the model could drive it down while reproducing sources poorly. A robust per-pixel loss is used, with additional weight on the bright cores of sources.

\subsection{From latent alignment to pixel reconstruction}

The masked-band objective was not the first design we tried, and the reconstruction approach was reached only after several alternatives failed to deliver the spatial precision the downstream tasks require. We summarize that history here because the negative results explain why reconstruction, rather than the more common alternatives, is the right objective for this problem.

The natural approach to relating the instruments is \emph{representation alignment}: encode a Rubin image and the matching \textit{Euclid} image each into a compact feature vector, and train the network so that the two vectors agree where they view the same sky. This is the principle behind contrastive learning \citep{contrastive} and Joint-Embedding Predictive Architectures (JEPA; \citealt{JEPA}), both widely used in computer vision. We designed and tested several cross-instrument variants in this family: patch-level contrastive matching, an object-centric model that attempts to discover individual sources, and JEPA architectures operating on the full image at native resolution with increasingly strict constraints forcing features at corresponding positions to match.

These approaches shared a common limitation that is structural rather than a matter of tuning. Alignment objectives reward features that are \emph{similar} but do not require the network to encode \emph{where} sources are located: two image regions can map to nearly identical feature vectors while differing in the sub-pixel positions of the sources they contain, and sub-pixel position is the quantity that astrometry and detection depend on. A direct control made this explicit: a simple supervised convolutional network, trained directly on the cross-instrument alignment task, outperformed the best self-supervised JEPA features at substantially lower cost.

Pixel reconstruction removes this failure mode by construction. To predict a withheld band at the correct positions, brightness, and morphology, the representation has no choice but to preserve the spatial layout of every source, a constraint that similarity-based objectives never impose. This is why the masked-band objective, rather than alignment, underlies the model adopted.

\subsection{Implementation}\label{sec:code}

The foundation model was trained on a single NVIDIA L40S GPU using \texttt{PyTorch} \citep{pytorch} with bfloat16 mixed-precision arithmetic. Optimization used AdamW \citep{adamw} with gradient clipping at unit norm, a five-epoch linear warmup followed by cosine decay, and an effective batch size of four, for 150 epochs; the checkpoint with the lowest validation loss was retained.

Pre-training completed in roughly 8 wall-clock hours ($\sim$8 GPU-hours). Once trained, the encoder is frozen and run once over the full dataset to extract and cache per-tile features; the fused detection head is then trained on these cached features. The astrometry head additionally reads raw VIS windows to compute its centroid anchor. The feature-extraction pass is the main per-dataset cost, but like the pre-training it is incurred only once and shared across all heads. Training was monitored with the Weights \& Biases platform.

The reconstruction loss uses a signal-and-gradient information map to emphasize source pixels while retaining a nonzero background weight. For a target image $I_p$ with per-pixel RMS $\sigma_p$, the reconstruction target is $t_p=50\,\operatorname{asinh}[I_p/(50\sigma_p)]$. This smooth compression preserves differences in bright-source cores. Writing $d_p=\hat{t}_p-t_p$, the implemented loss is
\begin{equation}
\mathcal{L}=s_{\sigma}\left[
\frac{1}{N}\sum_p w_p\sqrt{d_p^2+\epsilon^2}
+\lambda_{\rm core}\frac{\sum_p c_p d_p^2}{\max(1,\sum_p c_p)}
\right],
\label{eq:reconstruction_loss}
\end{equation}
where $w_p$ is the information weight normalized to unit mean, $c_p=\mathbf{1}(w_p>0.5)$ selects high-weight pixels, $\epsilon=10^{-3}$, and $\lambda_{\rm core}=0.2$. The scale factor is $s_{\sigma}=\max(\langle\sigma\rangle,0.1)$ in the input image units. The first term is a weighted Charbonnier penalty \citep{charbonnier1994}; the second adds squared-error sensitivity to source cores. The information map combines absolute pixel S/N and Sobel-gradient amplitude with an RMS-dependent background floor. Its full definition is given in the public code (Section~\ref{sec:release}).

\begin{figure}
    \centering
    \includegraphics[width=\columnwidth]{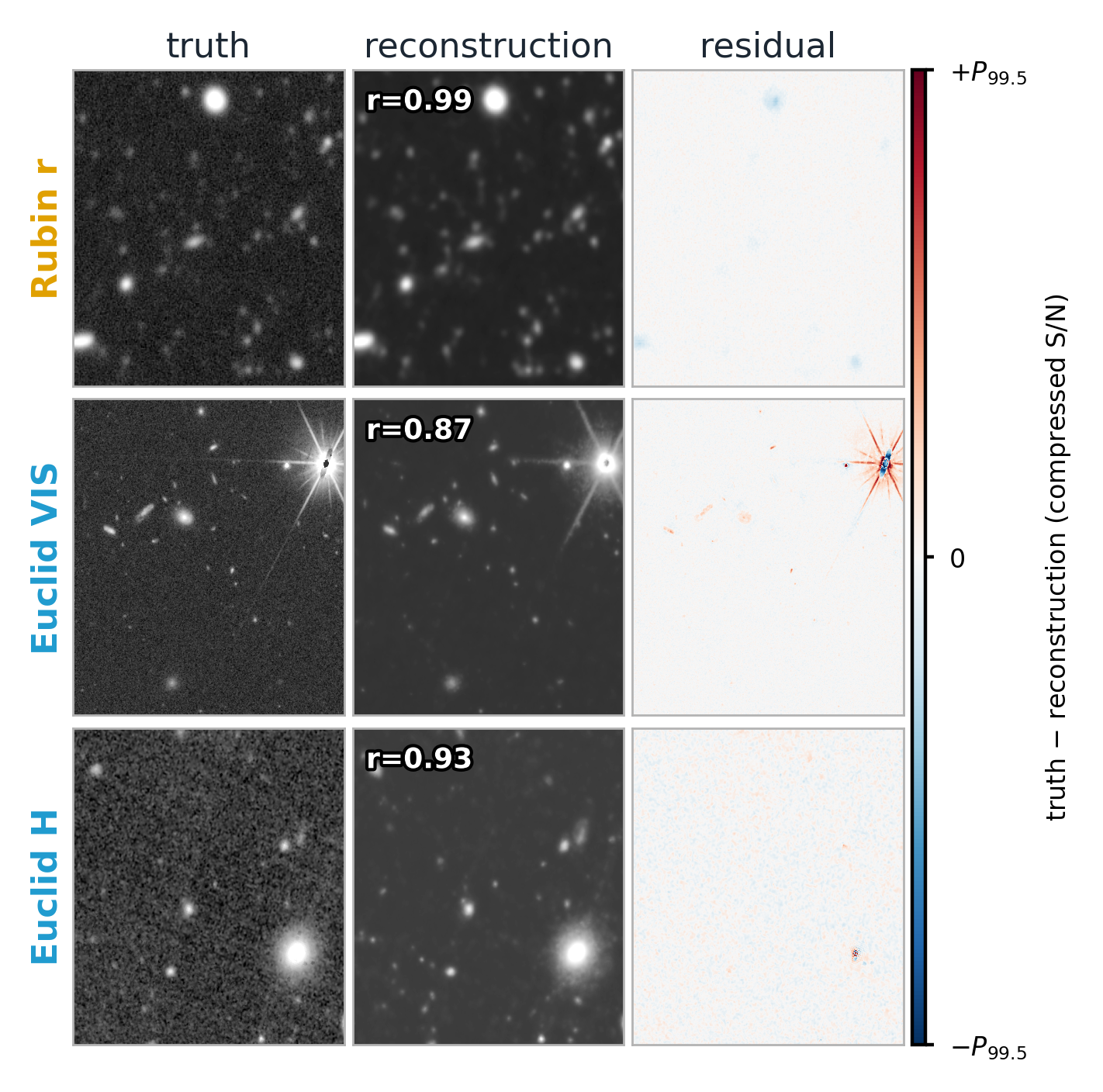}
\caption{Masked-band reconstruction on three held-out ECDFS tiles. In each row a single band is withheld and reconstructed from the other nine: Rubin $r$ (top), \textit{Euclid} VIS (middle), and \textit{Euclid} NISP $H$ (bottom). Columns show the true band (left), the reconstruction (center), and the residual truth $-$ reconstruction (right); panels are $48''$ on a side. Images use compressed S/N units, with residual limits $\pm P_{99.5}$ of each truth crop. After undoing the compression, the maximum absolute residuals are 45.8, 177.1, and 10.6 times the local input pixel RMS for Rubin $r$, VIS, and $H$, respectively; these extremes occur in the cores of the brightest point sources, where the compressed display saturates, and residuals elsewhere are far smaller. The annotated $r$ is the Pearson correlation between truth and reconstruction over the brightest 10\% of pixels; averaged across the 39 held-out tiles the mean per-band correlation is 0.99 ($r$), 0.77 (VIS), and 0.90 ($H$).}
    \label{fig:foundation_reconstruction}
\end{figure}

\subsection{Validation of the learned representation}\label{sec:validation}

Before attaching any task-specific head, we verify that the frozen representation has captured the cross-instrument structure it was trained to learn. We do this in three ways: by checking that held-out bands can be reconstructed faithfully, that the per-band information is recoverable from the frozen features a head would use, and that the model routes information between bands in a physically sensible way.

\textit{Reconstruction fidelity.} Figure~\ref{fig:foundation_reconstruction} shows masked-band reconstruction on validation tiles excluded from pretraining optimization, but potentially overlapping neighboring training tiles as described in Section~\ref{sec:data}. In each case a single band is withheld and predicted from the other nine, sampled across the full input range: a Rubin optical band, the high-resolution \textit{Euclid} VIS channel, and a \textit{Euclid} NISP near-infrared band are shown. The reconstructions reproduce the held-out sky faithfully, with residuals confined to the cores of the brightest point sources and the faint extended wings of large galaxies. Averaged over the held-out tiles, the truth-to-reconstruction correlation on the brightest tenth of pixels is $0.97$--$0.99$ for the deep, spectrally interior Rubin bands ($griz$), with the lower correlations plausibly reflecting the differences in noise, wavelength coverage, and resolution discussed below. The shallow bands $u$ and $y$ ($0.64$ and $0.84$) are limited by a noise ceiling: their truth images are themselves noisy, which bounds the correlation any prediction can achieve, and $u$ sits in addition at the blue edge of the ten-band range, where prediction becomes extrapolation. The NISP bands ($0.85$--$0.90$) draw chiefly on one another, as the attribution analysis below makes explicit. VIS, at $0.77$, is the one channel sharper than everything used to predict it: reconstructing it is an implicit deconvolution, asked to supply spatial frequencies the other nine bands do not contain, and the bright-pixel metric concentrates on exactly the point-source cores where those frequencies matter most. Some of this fidelity is expected on general grounds: galaxy spectral energy distributions are smooth, and for bright, well-detected sources the flux in a missing band can be recovered to first order by linear interpolation between its neighboring filters \citep{Hemmati2019}.
The reconstruction here carries more than that first-order transfer: it operates at every pixel rather than only on cataloged sources, it shapes its prediction to each source's morphology and surroundings instead of applying one global color law, and it crosses instruments, with their different pixel grids, samplings, and point-spread functions, so that a band from one telescope is predicted from the imaging of the other, something no within-instrument interpolation provides.

\begin{figure*}
    \centering
    \includegraphics[width=\textwidth]{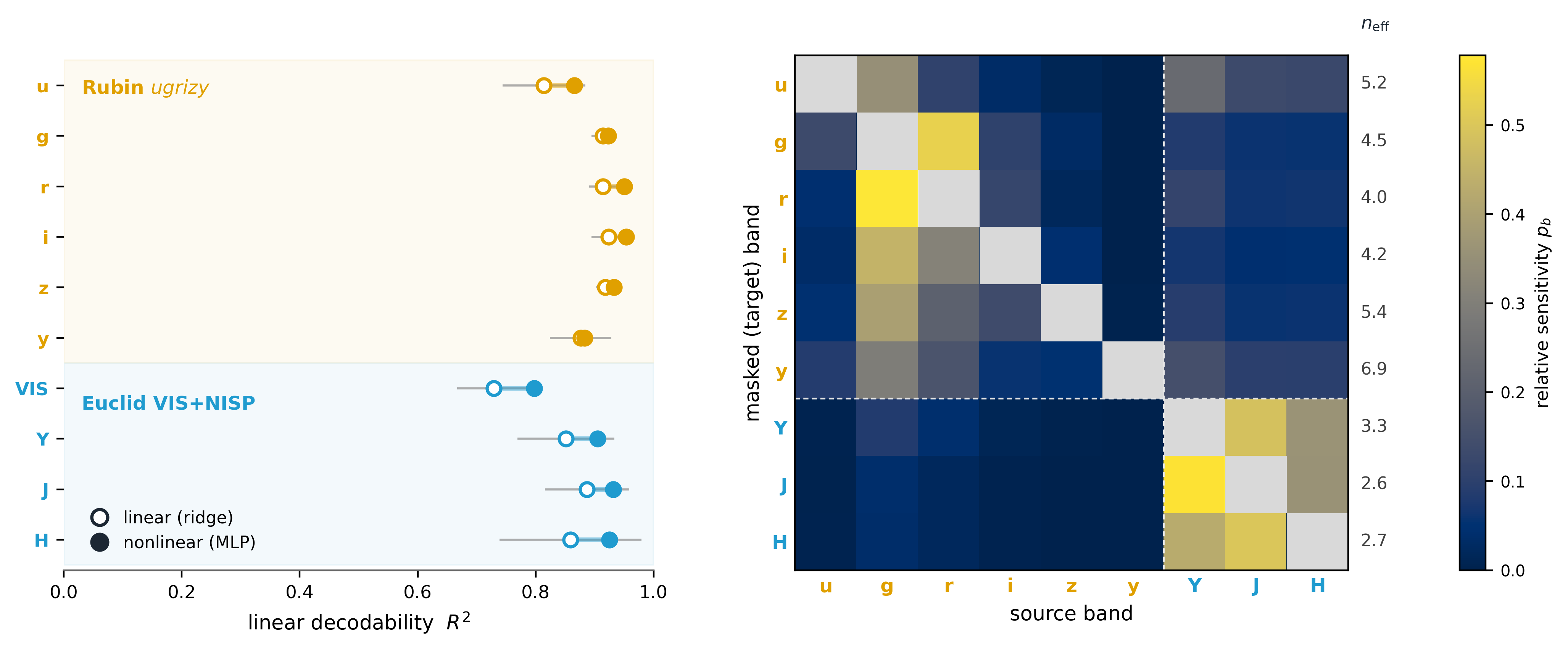}
\caption{Two views of the frozen foundation representation. \emph{Left:} prediction of each band's local $\log$\,S/N proxy from frozen features, evaluated with six folds grouped by tile (${\sim}770$ sources). Open and filled markers show linear (ridge) and nonlinear (MLP) probes; connecting lines show their difference, and gray bars show the standard deviation of the ridge scores across folds. \emph{Right:} relative sensitivity of the reconstruction loss to each input band (columns), for each withheld target band (rows). Colors show the normalized input-gradient weights $p_b$ defined in Equation~\ref{eq:attribution_sensitivity}; each row sums to one. VIS is excluded from the inputs to reveal relationships among the other bands, and gray diagonal cells mark the withheld targets. The effective number of contributing bands, $n_{\rm eff}=2^H$, uses the base-2 Shannon entropy of each row. Rubin reconstructions draw sensitivity from several bands, while NISP reconstructions depend mainly on the other NISP bands.}
    \label{fig:foundation_probe}
\end{figure*}

\textit{Decodability.} Faithful reconstruction is a statement about the network as a whole: the held-out band is produced by the encoder's features and the trained decoder acting together, so its success shows only that the information exists somewhere along that path, possibly folded into the decoder's weights. A downstream head inherits none of the decoder; it consumes the frozen bottleneck features alone. The question that matters for everything that follows is therefore whether those features, by themselves, hold the per-band information in a form a simple head can read out. We test this with a linear probe (left panel of Figure~\ref{fig:foundation_probe}): a ridge regression trained to predict each band's per-source local $\log$\,S/N proxy from the frozen features at the source position. Nothing learned enters the setup beyond the frozen encoder itself: the sources are simply $\mathrm{S/N}>5$ peaks in the smoothed VIS image, and the quantity to be recovered is $\log_{10}(\langle I\rangle/\langle\sigma\rangle)$ measured in a $5\times5$-pixel window at the source location in each delivered band. This is a local signal proxy rather than integrated-flux S/N with propagated aperture variance. The probe is evaluated with six folds grouping sources by tile, with features standardized per fold and the regression restricted per band to sources carrying signal in that band. These folds exclude test tile groups from probe fitting, but do not exclude their sky from encoder pretraining or guarantee independence of overlapping tiles. We quantify predictive performance with the coefficient of determination, $R^2=1-\sum_i(y_i-\hat y_i)^2/\sum_i(y_i-\bar y)^2$, where $y_i$ and $\hat y_i$ are the measured and predicted local $\log$\,S/N proxies, and $\bar y$ is the mean measured value in the test fold. A score of one indicates perfect prediction; zero corresponds to predicting the test-fold mean. All ten bands are recovered with high accuracy ($R^2 \approx 0.7$--$0.95$), and an exploratory nonlinear probe adds little, showing the per-band photometric information is not only present but \emph{linearly} available in the representation. We note that source brightness is an easy quantity to decode, so this is a check on representational richness rather than a strong claim about deep cross-instrument content; the more discriminating question is addressed next. The nonlinear comparison reports the best test-fold score over training epochs; it is therefore a diagnostic comparison rather than an independently selected test result.

\textit{Information routing.} The right panel of Figure~\ref{fig:foundation_probe} asks which bands the model actually draws on when reconstructing a given band. We measure this by input-gradient attribution, a standard interpretability technique whose idea is simple: the derivative of the reconstruction loss with respect to an input pixel measures the local sensitivity of that loss to a perturbation of the pixel. For each withheld target band, we quantify sensitivity to an input band $b$ using the mean absolute gradient of the reconstruction loss with respect to that band's pixels, divided by its mean absolute pixel value. We average this score over the evaluated tiles and normalize across the available input bands:
\begin{equation}
\begin{aligned}
S_b &= \left\langle
\frac{\left\langle\left|\partial\mathcal{L}/\partial I_b\right|\right\rangle_{\rm pixels}}
{\left\langle|I_b|\right\rangle_{\rm pixels}+\epsilon}
\right\rangle_{\rm tiles}, \\
p_b &= \frac{S_b}{\sum_{b'\in\mathcal B}S_{b'}}.
\end{aligned}
\label{eq:attribution_sensitivity}
\end{equation}
Here $I_b$ is the input image in band $b$, $\mathcal L$ is the reconstruction loss for the withheld target, $\mathcal B$ contains the available input bands, and $\epsilon=10^{-8}$ prevents division by zero. The weights $p_b$ sum to unity and describe relative local sensitivity under the adopted input scaling and loss; they do not measure fractions of reconstructed flux or causal contributions. The dominant result is that, when \textit{Euclid} VIS is available, it absorbs essentially all of the attribution: as the sharpest and deepest channel, it serves as a universal spatial scaffold on which every other band is reconstructed. Because this single-band dominance would otherwise mask all finer structure, the panel shown in Figure~\ref{fig:foundation_probe} is a counterfactual with VIS removed, exposing how the remaining nine bands route information among themselves. We summarize the distribution of these weights by the effective number of contributing bands, $n_{\rm eff}=2^H$, where $H=-\sum_{b\in\mathcal B}p_b\log_2p_b$ is the base-2 Shannon entropy, with zero-weight terms taken as zero. The effective number $n_{\rm eff}=2^H$ equals one when all sensitivity is concentrated in a single band and equals $k$ when it is distributed equally across $k$ bands. Unequal weights give an intermediate effective count. The Rubin optical bands have $n_{\rm eff} \approx 4$--$7$, drawing on many spectral neighbors with the adjacent bands contributing most, whereas the three NISP near-infrared bands form a near-closed block ($n_{\rm eff} \approx 3$) that relies chiefly on one another and barely on Rubin. Rubin $y$ contributes almost nothing to any reconstruction, including that of NISP $Y$, with which its bandpass overlaps: at a $5\sigma$ depth of $23.1$ it is the shallowest band in the stack, about a magnitude shallower than NISP $Y$, and the model draws instead on the deep $g$ and $r$ images, which dominate the Rubin inputs for every target. The tightness of the NISP block likely also reflects shared processing: the three bands are delivered on one MER grid through the same resampling, so they share spatial structure and correlated noise that no Rubin band does. This partly mirrors the structure of real galaxy spectral energy distributions, where neighboring bands are most predictive of one another, and is consistent with physically meaningful relationships between bands, although gradient sensitivity alone does not establish their origin.
Together these checks establish that the frozen representation is faithful, that its per-band information is linearly accessible, and that it routes information between bands in a physically coherent way. It is therefore a sound basis for the task-specific heads developed in the following sections.

\begin{table*}[!htbp]
\centering
\caption{Evaluation samples and their relation to training.}
\label{tab:evaluation}
\footnotesize
\setlength{\tabcolsep}{3.5pt}
\begin{tabular}{llll}
\toprule
Evaluation & Sample & Encoder exposure & Task-model training \\
\midrule
Reconstruction (Figure~\ref{fig:foundation_reconstruction}) & 39 ECDFS validation tiles & Neighboring sky may overlap & Decoder trained with encoder \\
Brightness prediction (Figure~\ref{fig:foundation_probe}) & ${\sim}770$ sources; 6 tile-group folds & ECDFS pretraining & Predictor trained on other folds \\
Detection (Figure~\ref{fig:detection_performance}) & 108 patch-25 tiles; & ECDFS pretraining & Patch 25 excluded \\
 & 45 for point-source completeness & & \\
Astrometry injections (Figure~\ref{fig:injection_truth}) & 108 patch-25 tiles & ECDFS pretraining & Patch 25 excluded \\
Astrometry maps (Figures~\ref{fig:astrom_agreement}, \ref{fig:head_correction}, \ref{fig:concordance_field}) & Full ECDFS footprint (790 tiles) & ECDFS pretraining & Includes evaluated sky \\
Field transfer (Figure~\ref{fig:ood}) & 72 EDF-S tiles & None & ECDFS only \\
\bottomrule
\end{tabular}
\end{table*}

\section{Detection}\label{sec:detection}
\begin{figure*}[t]
\centering
\includegraphics[width=\textwidth]{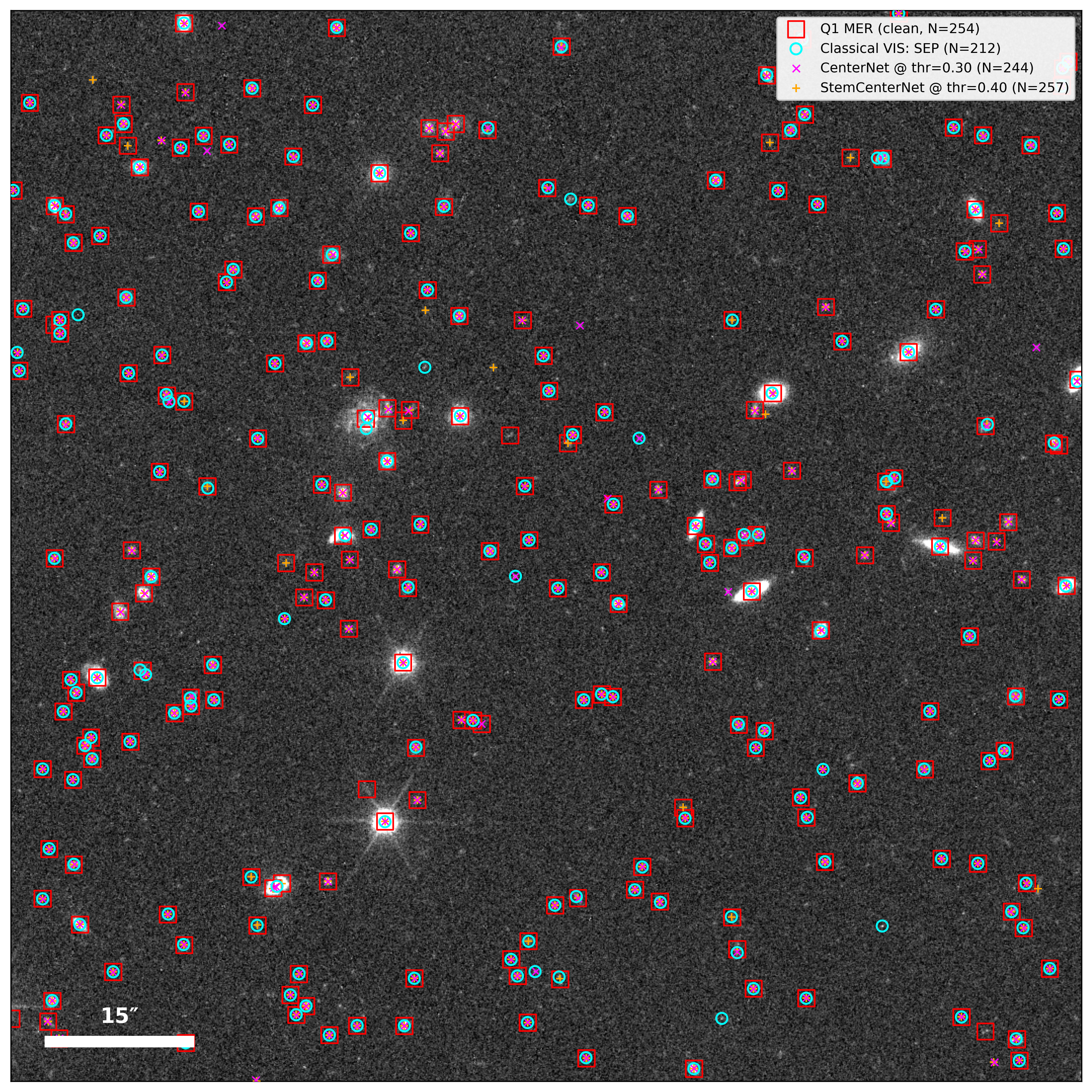}
\caption{Detection on a single held-out \textit{Euclid} VIS tile: the foundation head against classical and catalog references. A $1.7'$ tile (\texttt{tile\_x01792\_y00000}, $0.1\arcsecpx$, from a spatially disjoint patch excluded from detection training), with four source lists overlaid. \emph{Cyan circles:} the classical VIS source list (\textsc{SEP} extraction at $3\sigma$, cleaned of diffraction-spike artifacts and duplicates). \emph{Magenta crosses:} the foundation CenterNet head reading the frozen, fused ten-band bottleneck. \emph{Orange pluses:} the StemCenterNet variant reading the native per-instrument stems. \emph{Red squares:} the clean \textit{Euclid} Q1 MER catalog (VIS-detected, non-spurious), drawn larger so that the square frames the circle where both lists agree. The legend gives each list's count on this tile; a source with a cross but no cyan circle is a head detection below the \textsc{SEP} $3\sigma$ baseline, the depth gain quantified in Figure~\ref{fig:detection_performance}. The white bar is $15''$.}
\label{fig:detection_overlay}
\end{figure*}

\begin{table}[t]
\vspace{10pt}
\centering
\caption{Point-source detection depths.}
\label{tab:detection_heads}
\begin{tabular}{lcc}
\toprule
                       & CenterNet & StemCenterNet \\
                       & (fused)   & (stems)       \\
\midrule
$d_{50}$, all 10 bands & $26.7$  & $26.5$  \\
$d_{50}$, VIS only     & $26.4$  & $26.5$  \\
multi-band gain        & $+0.3$ & $+0.0$ \\
\bottomrule
\end{tabular}
\end{table}

Source detection is the first measurement we build on the frozen foundation, and it asks a specific question: does the frozen latent carry multi-band source evidence beyond what a single sharp band already provides, and can a lightweight head built on it match classical detection and then surpass it? Detection on \textit{Euclid} VIS alone is sharp and reliable, but it cannot use color or the Rubin and NISP channels in a learned way. We compare three source lists that differ only in what information they see: a classical VIS source-extraction list (\textsc{SEP}), which serves as the baseline and as the source of training labels, and two variants of a learned detection head built on the frozen foundation features. Comparing them isolates whether the representation has learned genuine multi-band source evidence, and whether that evidence is best expressed in the fused latent or in the higher-resolution per-band features.

\subsection{Heads and training labels}

Both variants share the same output design and differ only in the features they read. Following CenterNet \citep[][]{zhou2019}, each predicts a dense source-evidence map over the tile, in which a true source is represented during training by a small Gaussian peak at its position rather than by a single hard pixel, so that nearby pixels receive a graded target suited to sub-pixel positions. A focal loss \citep{lin2018focallossdenseobject} handles the extreme imbalance between the rare source pixels and the dominant empty sky. Alongside the evidence map each predicts a sub-pixel offset and a brightness proxy at every location. At inference the procedure is essentially learned peak-finding: local maxima above a confidence threshold are kept, and the offset is read off to refine each position. This is the same operation as classical peak-finding, performed on a learned ten-band source-evidence image. Classical pipelines also detect on multi-band combinations: the chi-squared image of \citet{Szalay1999}, the quadrature sum of per-band matched-filter images used for WISE \citep{MarshJarrett2012}, the per-band detection with cross-band merging of the HSC and Rubin pipelines \citep{Bosch2018}, and the VIS and NISP-stack detection of \textit{Euclid} MER \citep{MER}. Those combinations fix the band weighting in advance and assume matched PSFs; what the learned evidence map changes is that the weighting and the spatial combination across mismatched PSFs and samplings are learned from the data and vary with the local scene.

The two variants differ in where they read that evidence. The fused-bottleneck variant (CenterNet) operates on the frozen bottleneck, where the foundation model has already mixed all ten bands, and so tests whether the representation has learned multi-band source evidence. The native-stem variant (StemCenterNet) operates instead on the per-band stem features at their original resolution, projected onto a common grid, and so tests whether finer spatial detail helps localization beyond what the coarser bottleneck retains. Each has well under a million trainable parameters on top of the frozen encoder ($0.55$M for the fused variant): the encoder itself is never updated, and the heads read its precomputed feature maps, so the fused-head training can reuse cached features without rerunning the foundation.

We also tried a Detection Transformer \citep[DETR;][]{DETR}, which represents objects with a fixed set of learned ``query'' slots assigned to sources through a matching step. Its set-prediction machinery is built for the complexity of natural scenes and is data-hungry, and on our limited training set it converged slowly and unstably. A per-pixel design fits the problem better: every pixel receives a direct training signal, and detection reduces to finding peaks in a learned source-evidence image, much as sources are found classically. Blending, the main concern for such peak-based approaches, is also less severe here, since the fused representation brings the sharp \textit{Euclid} VIS scaffold to bear alongside the ground-based bands: the deconfusion of seeing-limited imaging by space-based resolution that joint processing was conceived to deliver \citep{Chary2020}, supplied here by the representation itself rather than by explicit positional priors, joint scene models \citep[e.g., \textsc{Scarlet};][]{Melchior2018}, or a dedicated deblending network \citep{Hemmati2022}.

We train the production head on labels generated directly from the VIS imaging using \textsc{SEP} \citep{Bertin1996,SEP} at $3\sigma$, followed by light cleaning: detections on diffraction-spike ridges are removed, each saturated star is consolidated to a single centroid, and near-duplicates are merged. We compared this extraction with an alternative label set from a multi-scale peak-finder: local maxima of the VIS image after Gaussian smoothing at four kernel widths from $0.9$ to $3$\,px ($0.09''$--$0.3''$), each thresholded against the robust background noise, then merged across scales and deduplicated within $0.5''$, so that the sharpest filter catches compact faint peaks while the broadest stabilizes the centers of extended galaxies. \textsc{SEP} produced the cleaner reference, with higher agreement with MER at matched purity.

\subsection{Reference catalog and evaluation design}

\begin{figure*}[t]
\centering
\includegraphics[width=\textwidth]{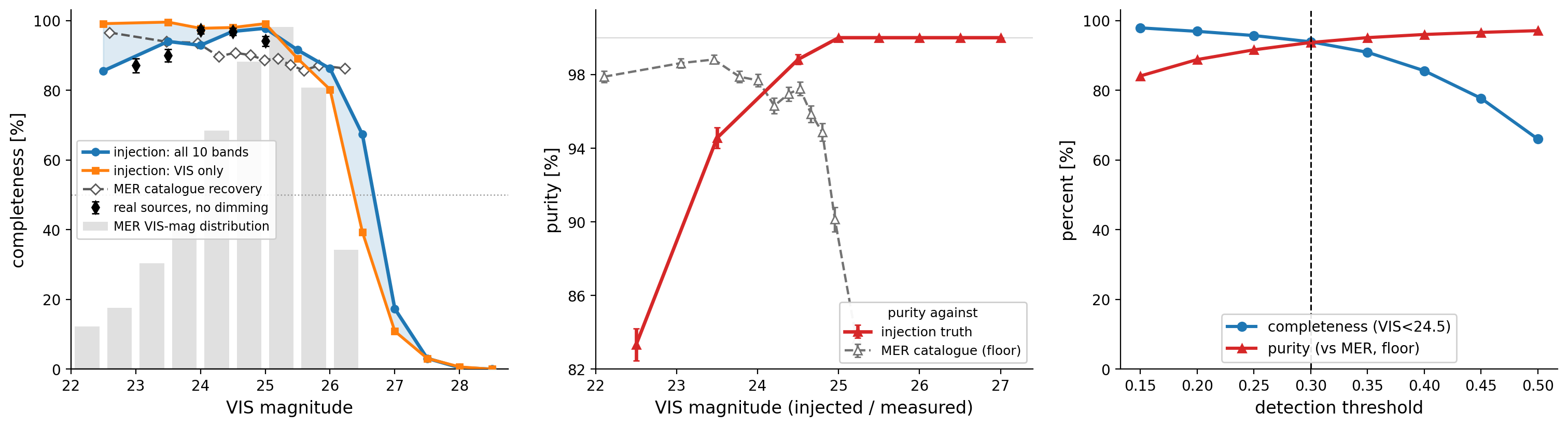}
\caption{Detection performance of the fused CenterNet head evaluated within the 108 patch-25 tiles excluded from head training. Point-source completeness uses the 45 tiles with sufficient local donors; the injection-purity test uses all 108 tiles. \emph{Left:} completeness versus VIS magnitude from source-recycling injection: real unresolved donors, carrying their full ten-band cutouts, dimmed to each target magnitude and placed in all ten bands (blue) or in VIS only (orange); the shaded band is the ${\approx}0.3$\,mag fusion gain, giving a $50\%$ point-source depth of VIS\,${\approx}26.7$. Black diamonds are real sources re-injected at native brightness, validating the dimming; their bright-end dip, shared with the all-band curve while in-situ MER recovery holds at ${\sim}97\%$, is an artifact of the copied stamps, not of the detector on real sky. The dashed curve is in-situ recovery of the clean VIS-selected Q1 MER sample and the gray histogram its VIS-magnitude distribution; a zero-flux control shows negligible chance recovery ($0$ of $450$ per mode). \emph{Center:} $P_{\rm inj}$ (Equation~\ref{eq:injection_purity}; filled, labeled ``purity against injection truth''), $100\%$ at the sampled magnitudes $25$--$27$ with the bright-end decline from bright-star wing detections, and catalog-based purity against the full MER catalog (open), whose faint-end drop tracks the MER turnover; it is marked ``floor'' here and in the right panel because MER incompleteness makes it a lower bound. The two statistics use different denominators. \emph{Right:} completeness (VIS\,$<24.5$) and catalog-based purity versus detection threshold; the dashed line marks the $0.30$ working point.}
\label{fig:detection_performance}
\end{figure*}

We evaluate detection performance in three complementary ways. First, we inject sources of known brightness and position to measure the fraction recovered as a function of magnitude, which defines the detection completeness and depth. Second, we examine whether adding these sources produces additional, unmatched detections in the images. Finally, we compare detections on the original sky with the published \textit{Euclid} catalog, measuring both recovery of catalog sources and the fraction of our detections with catalog counterparts. The injection-based purity describes changes caused by adding sources, while catalog-based purity measures agreement on the original sky.

We use the \textit{Euclid} Q1 MER catalog \citep{IRSA602} as an external cross-check. In Q1, \textsc{SourceExtractor++} detects sources independently in the VIS mosaic and a weighted stack of the three NISP bands, using two detection thresholds calibrated on simulations, followed by a dedicated density-based deblending stage. The lists are merged by retaining VIS detections and adding NIR detections outside the VIS segmentation regions \citep{MER}. For our catalog-completeness measurements, we select sources detected in VIS and not flagged as spurious (the catalog's \texttt{VIS\_DET} and \texttt{SPURIOUS\_FLAG} columns, respectively). This defines our clean VIS-selected reference sample; catalog-based purity instead uses counterparts in the full MER catalog.

As a control, we also trained the head on clean MER labels. The result is a slightly shallower detector: its $50\%$ point-source depth is $26.5$ in all bands and $26.2$ in VIS alone, about $0.2$\,mag below the production head, with a similar fusion gain of $+0.26$\,mag. Its agreement with MER is marginally higher, but that is not an independent validation for this control, since MER supplies its training labels. We retain the pixel-derived \textsc{SEP} labels for production because they reach fainter than the MER turnover, keep the stack free of external catalog supervision so that MER remains an independent check, and can be generated for any field with VIS imaging.

Figure~\ref{fig:detection_overlay} overlays the four source lists on a single held-out \textit{Euclid} VIS tile. The overall agreement is close: on the great majority of sources the four lists coincide. Two differences are visible on close inspection. First, the classical \textsc{SEP} list, even after the cleaning described above, tends to over-fragment extended galaxies, placing several detections across the spiral arms or star-forming clumps of a single resolved system, whereas the learned heads, though trained on those same labels, more often return one detection per object. Second, among the learned heads the fused CenterNet variant is the strongest: it recovers more faint sources than either the classical list or the native-stem variant, including objects that are faint in VIS but supported by the other bands. Whether the detector's added faint sources are real rather than noise is what the injection test below quantifies, by showing that real faint multi-band sources are in fact recoverable; the overlay shows the qualitative pattern, and the quantitative comparison follows.

\subsection{Injection recovery and detection depth}

Injection completeness is the fraction of added sources recovered at each magnitude: $C_{\rm inj}=N_{\rm recovered}/N_{\rm injected}$. We count a source as recovered when a new detection appears within $0.3''$ of its injected position. The $50\%$ detection depth, $d_{50}$, is the VIS-band magnitude at which half of the injected sources are recovered. This measures sensitivity beyond the reference catalog's faint-end completeness limit, where catalog matching alone cannot establish completeness.

We construct the injections by source recycling, a variant of the synthetic-source injection used to calibrate survey selection functions \citep[e.g., Balrog and SynPipe;][]{Suchyta2016, Everett2022, Huang2018}: real sources are selected and their image cutouts dimmed and inserted elsewhere in the images, so that the injected profiles carry the observed PSFs and colors rather than a model. We refer to these sources as donors. For the point-source test, donors are isolated, unresolved sources with $19.5<\mathrm{VIS}<22.5$, where VIS denotes magnitude measured in the \textit{Euclid} VIS band. Donors must have no cataloged neighbor within $3''$ and be compact in VIS, judged by aperture flux ratios in the spirit of the concentration indices used for morphological classification \citep{Abraham1994, Conselice2003}. All donors pass a loose cut, more than $60\%$ of the flux within a $1.1''$ box falling in its central $0.5''$; the point-source test adds a stricter one, $F(<0.3'')/F(<1.5'')\geq0.65$, which selects star-like profiles. The threshold was set empirically to exceed the median concentration of the real VIS $24$--$25$ population (${\approx}0.6$), so that dimmed donors are at least as compact as the sources whose depth they measure. This matters because the bright catalog contains many compact galaxies whose finite size would otherwise affect a point-source depth measurement. The procedure requires at least two qualifying donors within a tile. Of the 108 held-out patch-25 tiles, 45 meet this requirement, providing ten injections per magnitude per tile, or 450 injections per magnitude and band-selection mode.

We extract each donor's ten-band cutouts with a radius of $3''$, subtract a local background estimated from each cutout's edges, and dim them to a chosen target magnitude. This retains the observed PSFs and spectral energy distribution within the cutouts. We place them at positions with valid VIS coverage at least $1''$ from any source in the clean MER reference and $0.8''$ from any pre-injection detection, then rerun the frozen detector. Recovery is scored against the known injection positions rather than the catalog, while the magnitude scale is inherited from the donors' bright-end MER photometry.

To measure the benefit of combining bands, we compare injections in all ten bands with injections in VIS alone. The detector receives the full ten-band scene in both cases; only the bands carrying the added source change. The fused head reaches a $50\%$ point-source depth of VIS magnitude 26.7 for all-band injections and 26.4 for VIS-only injections, a gain of approximately 0.3 magnitudes (Figure~\ref{fig:detection_performance}, left). The identical comparison for StemCenterNet gives no measured gain from the other nine bands ($+0.0$\,mag versus $+0.3$\,mag; Table~\ref{tab:detection_heads}). Thus the gain in this comparison is associated with the fused-feature head, which we adopt for subsequent analyses.

Repeating the test with compact-galaxy donors gives depths of 25.2 and 24.7 for all-band and VIS-only injections, respectively. The shallower depths are expected when the light is spread over a larger area, but the 0.5-magnitude gain shows that the multi-band advantage also occurs for moderately extended profiles. Injecting into NISP alone gives only marginal recovery, while injecting into Rubin alone gives essentially none. At these depths, the head remains VIS-anchored: the other bands add sensitivity in combination with VIS but rarely drive detections independently.

Two controls validate the procedure. First, real sources re-injected at their native brightness, without dimming, are recovered at the same rate as dimmed donors at VIS 23--25, confirming that the dimming step does not bias the completeness curve (black diamonds in Figure~\ref{fig:detection_performance}). Second, a zero-flux control measures how often a recovery is scored when no source light is added, for instance when a noise peak happens to fall within the $0.3''$ matching radius: each donor is dimmed to VIS 35 and processed identically. None of the 450 injections per mode is recovered, so chance recovery under this protocol is negligible, and the handful of recoveries at VIS 28 and fainter is consistent with zero.

\subsection{Injection-based purity from detection-list differencing}

Adding a source can also produce extra detections, for example on its wings. We measure injection-based purity by differencing the detection lists before and after injection: detections already present in the original scene are matched and excluded, leaving the newly appearing detections to compare with the known injections. This is a comparison of source lists, rather than subtraction of the images. Newly appearing detections can correspond to the injected source, to a pre-existing faint peak pushed above the detection threshold, or to an additional detection with neither counterpart. We count these categories separately so that recovery of an injected source is distinguished from other changes in the scene.

For this test we draw donors from a shared library assembled across ECDFS, with repeated sources removed and no cataloged neighbor within $7''$. This allows all 108 held-out tiles to receive 15 injections per magnitude, giving 1620 injections per magnitude and mode. We use cutouts of radius $6''$ with smoothly tapered (apodized) edges, so that truncating the donor image is less likely to create artificial detections. The wider donor-isolation cut reduces contamination by neighboring sources copied with the donor. Injected centers are kept far enough from tile edges to contain the full cutout, and are separated from one another by at least $2''$. Only one magnitude is injected per detector pass, allowing the resulting changes to be associated with that injection magnitude. We compare detection lists at the adopted score threshold of 0.30 and also record pre-injection peaks down to a score of 0.02 to identify faint peaks promoted above the working threshold.

Classification uses two radii. An injection match uses the same $0.3''$ radius as the completeness test. A detection is counted as newly appearing only if it lies more than $0.4''$ from every pre-injection detection at the working threshold; the slightly larger radius absorbs the small centroid shifts that added flux can induce in existing detections, so that a displaced re-detection of an old source is not mistaken for a new one. Among the newly appearing detections, those within $0.3''$ of a peak in the low-threshold pre-injection list are classified as promoted pre-existing peaks; the remainder are unmatched induced detections: false positives attributable to the injection itself, such as secondary peaks on the injected source's wings.

The plotted statistic is
\begin{equation}
P_{\rm inj}=\frac{N_{\rm recovered}}{N_{\rm recovered}+N_{\rm induced}},
\label{eq:injection_purity}
\end{equation}
where $N_{\rm recovered}$ is the number of recovered injections and $N_{\rm induced}$ is the number of unmatched induced detections. Promoted pre-existing peaks are reported separately and excluded from this denominator. Thus $P_{\rm inj}$ measures the balance between recovered injections and unmatched induced detections. It does not measure the purity of the full output catalog, since the underlying real sky is not a complete truth catalog.

No unmatched induced detections occur at the sampled VIS magnitudes 25--27 in the all-band test, giving an observed $P_{\rm inj}=100\%$ there (Figure~\ref{fig:detection_performance}, center). At VIS magnitude 22.5, secondary detections on bright-source wings reduce this statistic to $83\%$, compared with approximately $90\%$ for VIS-only injections at the same magnitude. For example, at VIS magnitude 26.5 the all-band test recovers 1242 of 1620 injections, with zero unmatched induced detections and 146 promoted pre-existing peaks. The latter are excluded from $P_{\rm inj}$, giving $100\%$ for this statistic even though some faint peaks cross the detection threshold. These finite-sample results constrain unmatched induced detections under the stated classification; they do not establish a zero false-positive rate on arbitrary sky, where no truth catalog exists.

\subsection{Catalog comparison and limitations}

The MER catalog provides a complementary test on the original sky, where real sources retain their observed neighbors and blending. It is not fully independent of our training labels, since MER detection is itself \textsc{SourceExtractor}-based, as are the \textsc{SEP} labels, although the two differ in thresholds, deblending, and the NIR-stack detection stage. Catalog completeness is the fraction of sources in the clean VIS-detected reference sample recovered by our detector. Catalog-based purity is the fraction of our detections with counterparts in the full MER catalog. Both catalog comparisons use a $0.5''$ matching radius, looser than the $0.3''$ used against injected truth because MER centroids follow a different convention from our peak-centered positions, sitting off the peak for extended or blended sources. The former asks how much of the reference sample we recover; the latter asks how many of our detections the reference can account for.

The detector recovers approximately 85--97\% of the clean reference across the magnitude range reached by MER (Figure~\ref{fig:detection_performance}, left, dashed curve). Raising the detection threshold increases catalog-based purity while reducing completeness (Figure~\ref{fig:detection_performance}, right panel); we adopt a working threshold of 0.30.

An unmatched detection need not be false, because MER has its own selection function and becomes incomplete at faint magnitudes. Unmatched candidates include sources with Rubin or NISP aperture flux above that at random sky positions, supporting a contribution from real multi-band sources absent from MER. Catalog incompleteness can therefore reduce the catalog-match fraction. A classical multi-band detection image built from the same ten bands, in the manner of \citet{Szalay1999}, would be the natural intermediate baseline; we have not constructed one here, since the mismatched PSFs would first require homogenization, and the fusion gain reported above is therefore measured against VIS alone. Catalog-based purity and $P_{\rm inj}$ answer different questions and use different denominators: one describes agreement with an external catalog on the original sky, and the other counts unmatched induced detections from adding known sources.

In summary, the head begins from classical VIS detection and extends it: it is deeper, uses all ten bands, and shows less fragmentation of extended systems in the examples examined.

\section{Astrometry}\label{sec:astrometry}

\begin{figure*}[t]
    \centering
    \includegraphics[width=\textwidth]{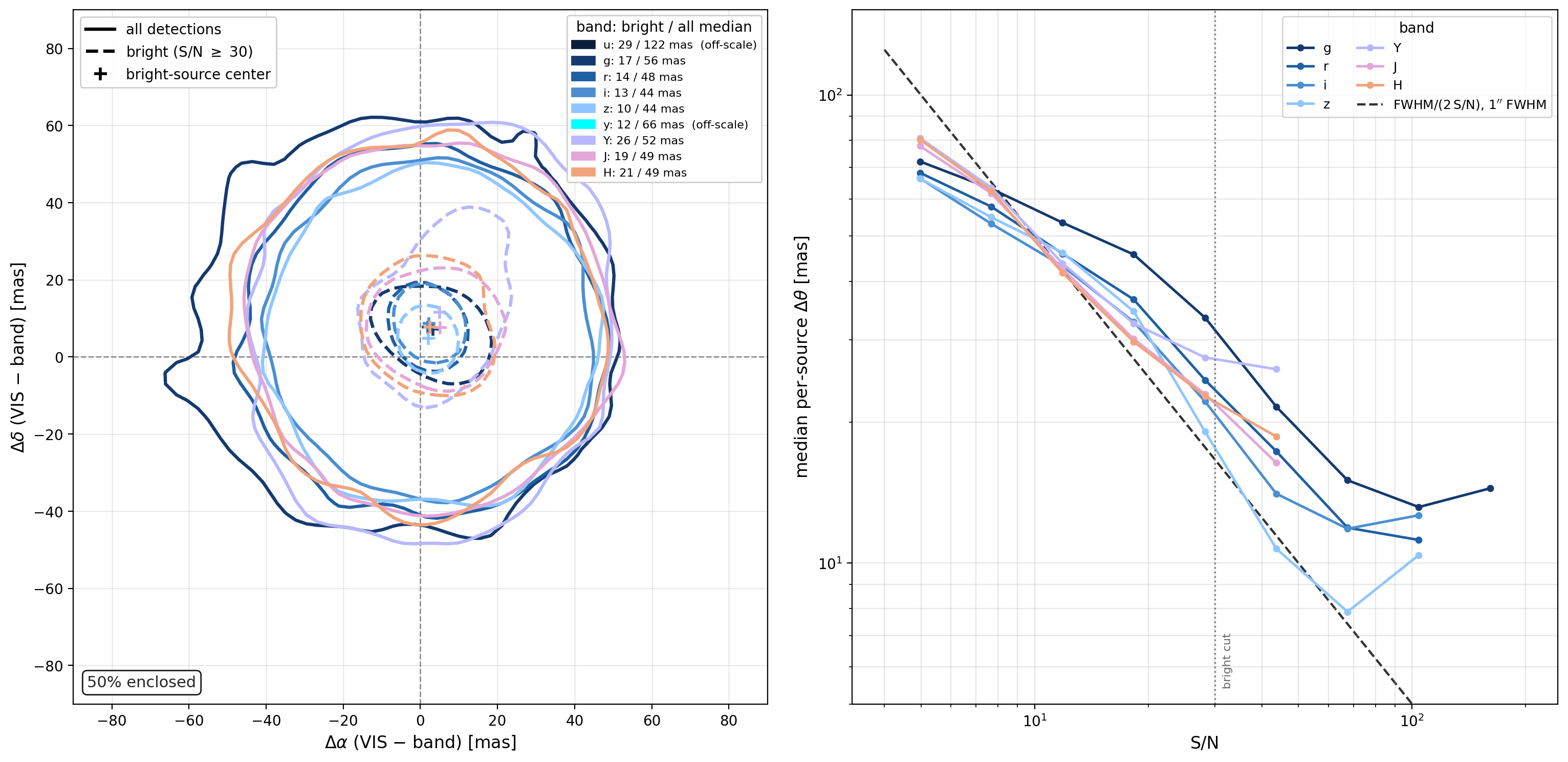}
\caption{Raw astrometric agreement between Rubin/NISP and \textit{Euclid} VIS before any correction, from classical Gaussian-centroid positions of matched sources pooled over the ECDFS field; plotted offsets are $(\Delta\alpha,\Delta\delta)$ as defined in Equation~\ref{eq:offsets}, i.e.\ VIS $-$ Rubin/NISP. Each band's sample is the set of sources detected in that band and matched to VIS; detection in the other bands is not required, so the per-band samples differ in size. \emph{Left:} two-dimensional offset clouds, each contour enclosing $50\%$ of that band's sources; solid contours are all detections, dashed contours bright sources ($\mathrm{S/N}\geq30$). Colored crosses mark the bright-source centers, defined by the coordinate-wise median offsets. Although both surveys are independently registered to \textit{Gaia}, the bright clouds do not close on the origin: their centers are displaced by ${\sim}7$--$10$\,mas, most of it a common shift shared across all ten bands, coherent but not constant over the footprint (the concordance field mapped in Figure~\ref{fig:concordance_field}). The NISP cloud centers sit slightly farther (${\sim}10$\,mas) than the Rubin ones (${\sim}7$\,mas; see text). The legend lists the bright/all median per-source offset per band; the $u$ and $y$ clouds lie off the plotted scale and are given as numbers only. \emph{Right:} median per-source radial offset versus the source's S/N in that band, for the seven well-measured bands; the dashed line is the photon-noise scaling of centroid precision, $\mathrm{FWHM}/(2\,\mathrm{S/N})$ for a $1''$ FWHM \citep{king1983}, drawn as a guide (see text). The dotted line marks the $\mathrm{S/N}=30$ bright cut.}
    \label{fig:astrom_agreement}
\end{figure*}

\begin{figure*}[t]
    \centering
    \includegraphics[width=\textwidth]{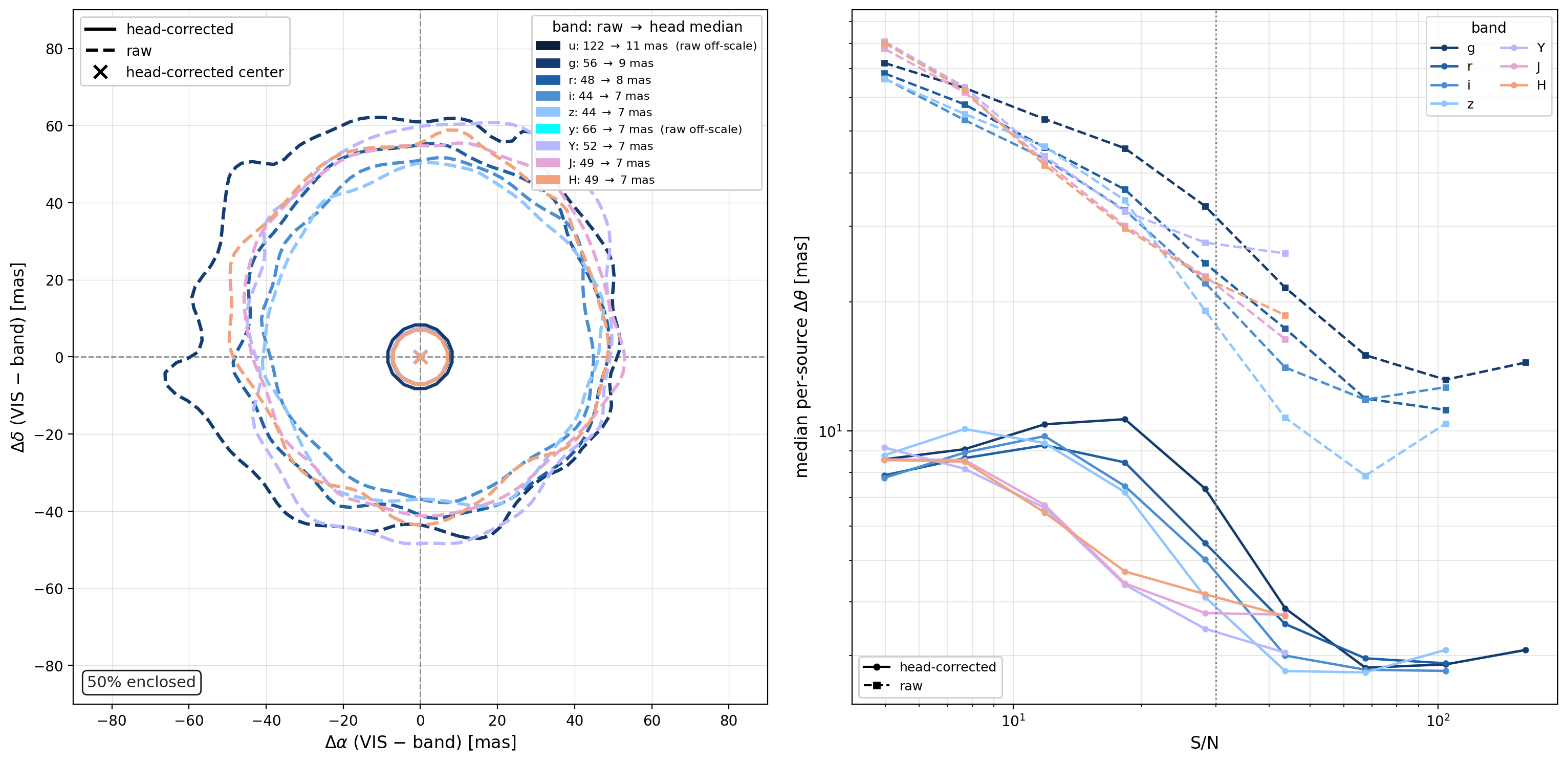}
\caption{Effect of the per-source astrometry head on cross-survey agreement, using the same offset measurement as Figure~\ref{fig:astrom_agreement} (\textit{Euclid} VIS $-$ Rubin/NISP; Equation~\ref{eq:offsets}) over all 790 ECDFS tiles, including head-training sky. Each band's sample is the set of sources detected in that band and matched to VIS; detection in the other bands is not required, so the per-band samples differ in size. Solid curves are head-corrected positions; dashed curves are the raw classical centroids. \emph{Left:} all-detection offset clouds, each contour enclosing $50\%$ of that band's sources. Colored $\times$ markers show the coordinate-wise median centers of the head-corrected clouds. The head collapses the raw ${\sim}50$\,mas clouds to ${\sim}6$--$10$\,mas and re-centers them on the origin, removing both the per-source scatter and the systematic offset; the legend gives the raw$\,\rightarrow\,$head median per band ($u$ and $y$ raw clouds lie off the plotted scale). \emph{Right:} median per-source radial offset versus the source's S/N in that band: the head-corrected curves (solid) are far flatter than the raw (dashed), staying near the bright-source floor across the full range, with the in-model centroid anchor holding the bright end at the classical systematics floor. These medians measure agreement with the VIS-centroid convention that also supplies the head's training labels, so they can sit below the accuracy against truth, which is bounded by the VIS anchor and measured directly in Figure~\ref{fig:injection_truth}.}
    \label{fig:head_correction}
\end{figure*}

\begin{figure}[t]
    \centering
    \includegraphics[width=\columnwidth]{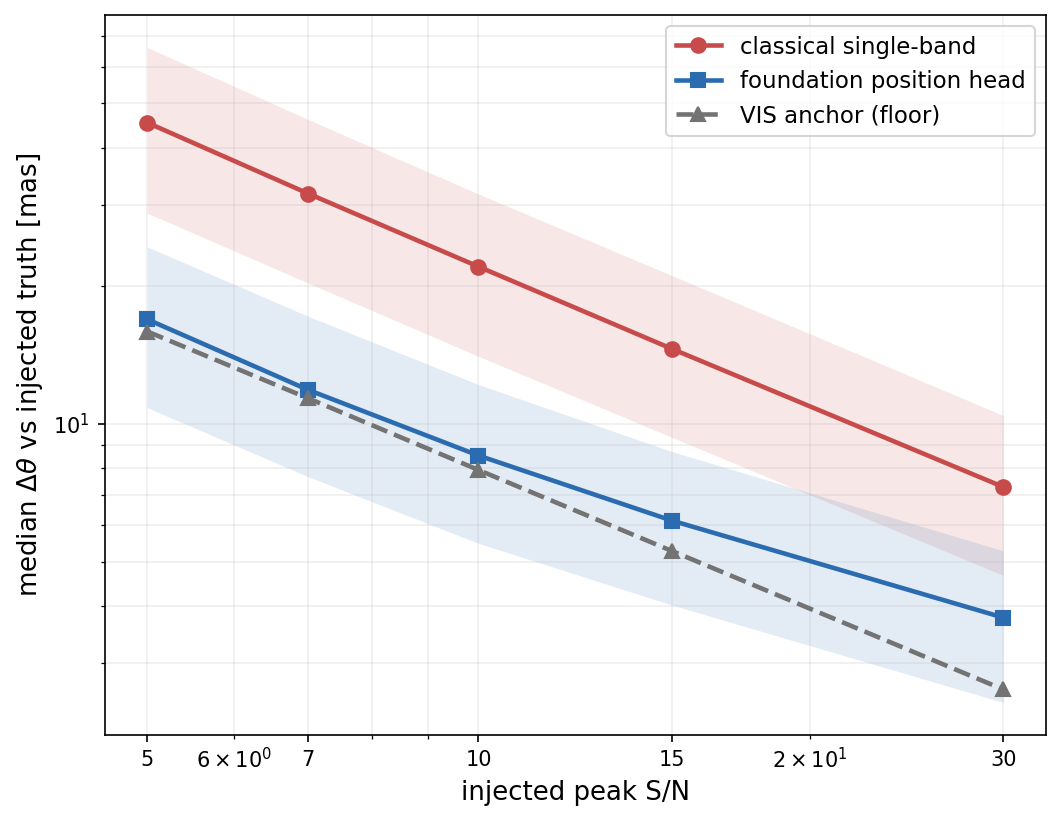}
\caption{Injection-vs-truth astrometry: the astrometry head transfers VIS-level localization to the Rubin bands. Sources of known position are injected into all ten bands of held-out tiles at controlled peak S/N; the median radial error is measured against the \emph{injected truth} (not the VIS label) and pooled over the Rubin optical bands ($griz$), with shaded 25--75\% ranges. The classical single-band Gaussian centroid (red) rises toward faint sources as expected for photon-noise-limited centroids, reaching ${\sim}45$\,mas at $\mathrm{S/N}=5$. The foundation astrometry head (blue) instead tracks the VIS anchor error (gray, dashed), the floor set by the VIS centroid the head is trained to reproduce, collapsing the $\mathrm{S/N}=5$ error to ${\sim}17$\,mas, within ${\sim}1$\,mas of the ${\sim}16$\,mas VIS floor.}
    \label{fig:injection_truth}
\end{figure}

\begin{figure*}[t]
    \centering
    \includegraphics[width=\textwidth]{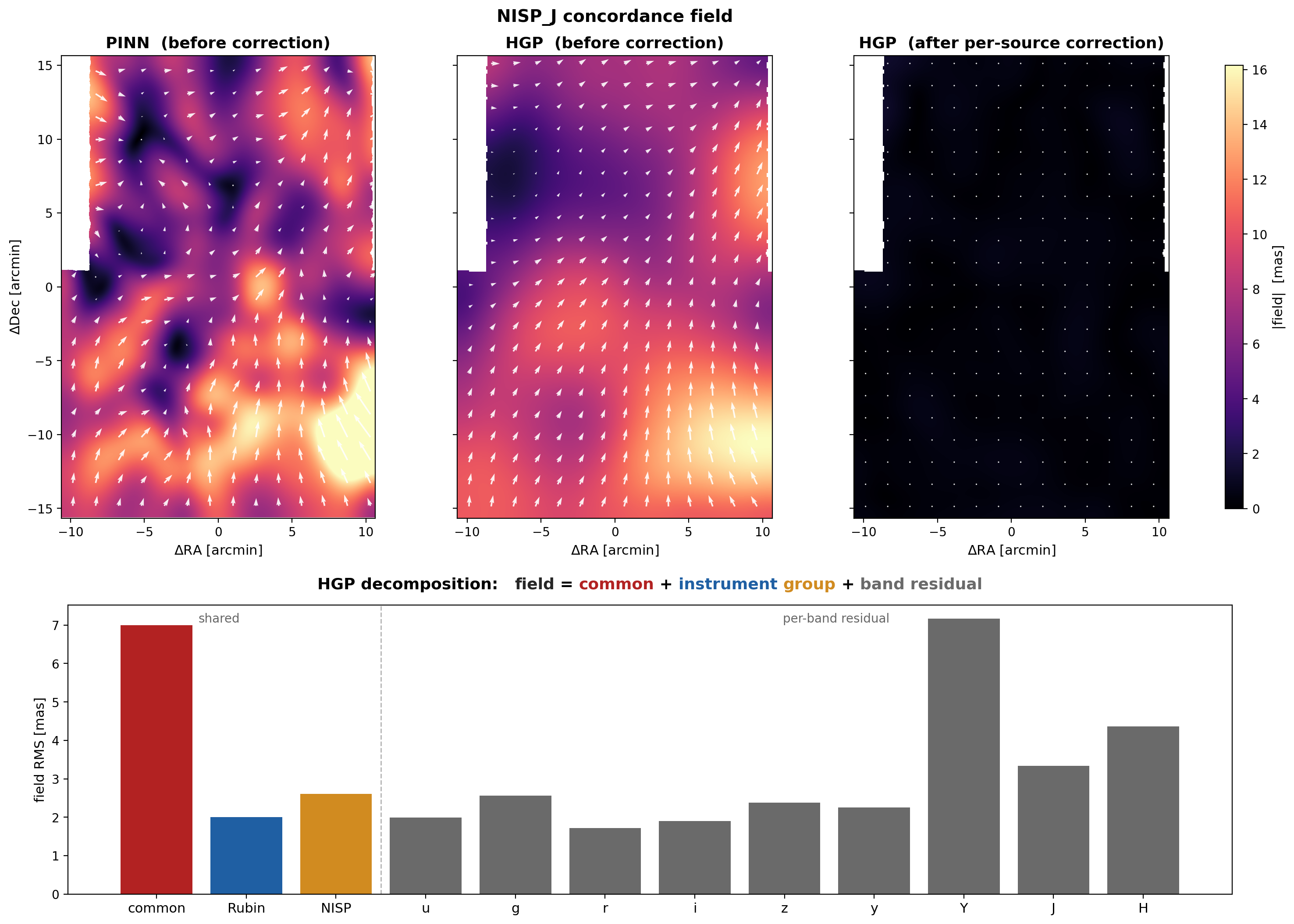}
\caption{The cross-survey concordance field before and after the per-source correction. \emph{Top:} the NISP\,$J$ residual field over the ECDFS footprint (arrows: direction; color: $|$field$|$; smoothed to $60''$). Before any correction, the field is measured with two solvers: the physics-informed network (PINN, left) and the hierarchical Gaussian process (HGP, center). They recover the same ${\sim}9$--$10$\,mas pattern, the local amplitude drifting smoothly between ${\sim}5$ and ${\sim}13$\,mas across the footprint. Refit to the head-corrected offsets (right, same color scale), the field amplitude drops to ${\sim}1$\,mas. \emph{Bottom:} the HGP hierarchical decomposition of the pre-correction field into a component common to all bands (${\sim}7$\,mas), per-instrument group terms (Rubin, NISP), and small per-band residuals ($u$--$H$); the dominant common term shows most of the field is a single shared cross-survey systematic rather than per-band noise.}
    \label{fig:concordance_field}
\end{figure*}

Astrometry is the second measurement drawn from the frozen foundation. Rubin and \textit{Euclid} each deliver a mature astrometric solution, independently registered to the Gaia reference frame \citep{gaia2016, gaia2023}, and we do not attempt to improve either. Those delivered WCS solutions define the coordinate maps used to compare bands, but no external catalog positions enter the learned per-source correction or the concordance-field fit: both are built from the pixels alone, from where the instruments see the same sources. The common Gaia tie also sets up a stringent test: two independently registered solutions should already agree, and whatever disagreement remains is a systematic that any joint measurement would inherit.

This section develops two products: a per-source astrometry head that transfers the sharp VIS localization to the nine other bands, where the PSF is broader or the source fainter, and a measurement of the residual agreement between the two WCS solutions, which we call the cross-survey concordance field. The name is inherited from the concordance frame of \citet{Chary2020}: cross-survey astrometric agreement at the ${\sim}10$\,mas level, matching each survey's internal precision, was the threshold that study set for joint measurement to proceed, and the field mapped below is a first measurement of where two independently calibrated solutions actually stand against it. The two products live at very different amplitudes, and separating them cleanly is the main work of this section.

\subsection{Positional offsets and reference conventions}

Throughout this section an astrometric offset is a per-source positional difference between two measurements of the same object. For a source $i$ detected in a band $b$ and matched to its VIS counterpart, we define
\begin{equation}
\Delta\alpha_i = \left(\alpha_{\mathrm{VIS},i} - \alpha_{b,i}\right)\cos\delta_i , \qquad
\Delta\delta_i = \delta_{\mathrm{VIS},i} - \delta_{b,i} ,
\label{eq:offsets}
\end{equation}
where $(\alpha,\delta)$ are the sky positions obtained from each mosaic's delivered WCS and the factor $\cos\delta_i$, evaluated at the VIS declination, converts the RA difference to a distance on the sky. The sign convention is VIS minus the other band, so positive values mean the VIS position lies at larger RA or Dec than the band position. From these we form the radial offset $\Delta\theta_i = (\Delta\alpha_i^2 + \Delta\delta_i^2)^{1/2}$ and use two summary statistics. The first is the median of $\Delta\theta_i$ over a set of matched sources, which we call the median per-source offset; it measures cross-band positional agreement, including random and systematic contributions. The second is the displacement of the center of the offset cloud from the origin, $(\widetilde{\Delta\alpha}, \widetilde{\Delta\delta})$, where the tilde denotes the median over the sample; its magnitude measures a coherent systematic between the two astrometric solutions, with uncertainty depending on the source sample and positional scatter. We use medians rather than means and RMS values because the faint majority of matched sources has heavy-tailed centroid errors; for Gaussian scatter with per-axis dispersion $\sigma$ and no systematic, the median radial offset is $1.18\,\sigma$. The astrometry head described below predicts, for each source and band, a correction to the classical band position; adding it gives the head-corrected band position, and substituting that for $(\alpha_{b,i},\delta_{b,i})$ in Equation~\ref{eq:offsets} gives the head-corrected offsets.

Figure~\ref{fig:astrom_agreement} shows the starting point: the raw astrometric agreement between the surveys before any correction, from classical Gaussian-fit centroids of sources matched between each band and VIS, pooled over the training field ($0.4$--$2.7\times10^{4}$ matched pairs per band, ${\sim}1.5\times10^{5}$ in total). Bright sources (S/N $\geq 30$; hundreds to thousands per band) already agree well, with median per-source offsets of ${\sim}10$--$25$\,mas per band, but their offset clouds do not close on the origin: the center of each cloud is displaced from zero by ${\sim}7$--$10$\,mas, a coherent systematic distinct from the per-source scatter, most of it a common shift shared across all bands (the concordance field measured below). The NISP cloud centers sit slightly farther from the origin (${\sim}10$\,mas) than the Rubin ones (${\sim}7$\,mas) despite the NISP mosaics being delivered on the same pixel grid as VIS; a plausible origin for the small excess is the extra processing the NISP imaging undergoes before reaching that grid, the registration and resampling of the native $0.3''$ frames onto the $0.1''$ MER mosaics (Section~\ref{sec:data}), though the mosaics alone do not let us isolate the cause. Persistence in the NISP detectors is another candidate. Q1 masks only the persistence carried over from the preceding spectroscopic exposure, and the release notes that unmasked residual streaks remain and that persistence between consecutive photometric exposures is not treated \citep{NIR}, and such residuals can bias centroids.

The full source population, dominated by faint galaxies, behaves differently: its median per-source offset is ${\sim}45$--$55$\,mas, and the right panel shows that this disagreement follows the expected photon-noise scaling of centroid precision with FWHM and S/N \citep{king1983}, drawn there as the dashed guide. The guide's normalization is indicative rather than exact: it adopts a single $1''$ FWHM for all bands, the aperture S/N on the horizontal axis understates the total-photon S/N of the underlying derivation, and the faintest bins are trimmed by selection, since a source must be detected and matched to enter the sample at all. These conventions let the Rubin curves sit slightly below the line without any centroid outperforming the physical limit, while the NISP curves sit above the lower expectation their sharper PSF would set, plausibly reflecting the undersampled native $0.3''$ imaging and the correlated noise introduced when it is resampled onto the MER grid.

What the comparison establishes is the slope: all bands follow the ${\sim}1/(\mathrm{S/N})$ decline through the faint regime, identifying the disagreement of the full population as photon noise rather than a calibration failure, before flattening onto the bright-source floor, set by the cloud-center displacement and the VIS centroid error, which photons cannot reduce. The raw offset is therefore not one effect but two, different in nature and very different in amplitude.

This decomposition dictates the strategy. The coherent component is small, ${\sim}7$--$10$\,mas, and is carried by bright sources that are sparse on the sky (${\sim}1$--$3\,\mathrm{arcmin}^{-2}$); it is a smooth field to be mapped. The dominant component is per-source centering noise: it lives at the photon-noise floor set by each band's PSF and depth, it is uncorrelated between neighboring sources, and no global correction can remove it. It has to be corrected one source at a time, conditioned on the actual image content, which is precisely what a head on the foundation representation can do.

\subsection{The astrometry head and its VIS anchor}

The astrometry head is a small network (${\sim}0.7$M parameters) attached to the frozen encoder. For each detected source it reads two local windows: one from the fused bottleneck, whose $0.4\arcsecpx$ features carry the ten-band morphology, and one from the VIS stem features at their native $0.1\arcsecpx$ sampling, which provide the fine spatial reference. From these it predicts a per-source, per-band position offset together with its uncertainty. The prediction is tied to a classical centroid: in its forward pass the head computes an iterated Gaussian-weighted centroid on a $17\times17$ raw VIS window at the query position, together with that centroid's photon-noise uncertainty from the same pixels, and the network adds only a residual bounded per coordinate at three times the estimated centroid uncertainty, with implemented bounds of 5--250\,mas. For bright compact sources the centroid uncertainty is a few mas and the network can change the position only slightly; for faint sources the uncertainty, and hence the bound, is larger, and the learned residual contributes most of the correction. Training is self-supervised and requires no external catalog: VIS Gaussian-fit centroids serve as targets, the query position is perturbed by a controlled jitter, and the head is trained to recover the centroid from the local features. The loss is a Rayleigh negative log-likelihood with a fixed label-noise floor of 5\,mas, reflecting the systematic floor of the underlying WCS solutions, so that the head is not penalized for residuals below the noise in individual bright-source labels. The predicted uncertainty weights this likelihood, down-weighting ambiguous sources during training; its calibration is tested below.

\subsection{Validation against VIS, injections, and Gaia}

Figure~\ref{fig:head_correction} shows the effect. The head collapses the raw ${\sim}50$\,mas offset clouds to ${\sim}6$--$10$\,mas medians in every non-VIS band and re-centers them on the origin, removing the per-source scatter and the coherent band offsets together. The S/N dependence flattens correspondingly: the median offset of faint sources decreases to near the bright-source value, removing most of the $1/(\mathrm{S/N})$ scaling of the single-band centroids in Figure~\ref{fig:astrom_agreement}, although a weak residual dependence on S/N remains. Because the head is trained against classical VIS centroids, agreement with those labels could in principle reflect emulation of the label convention rather than genuinely better positions; we test this directly by injecting sources of known position into all ten bands of the 108 patch-25 tiles (Figure~\ref{fig:injection_truth}). The plotted head excludes patch 25 from head training, while its encoder retains ECDFS pretraining. The injections assign positions through the delivered WCS, testing measurement errors conditional on those mappings rather than absolute WCS accuracy. At a peak S/N of 5, head positions err against truth by ${\sim}17$\,mas, compared to ${\sim}45$\,mas for classical single-band centroids, close to the ${\sim}16$\,mas floor set by the VIS anchor itself. This is the transfer the section set out to test: the head carries the VIS localization to the other bands, and adds no precision beyond what VIS provides.

Two conventions govern how these numbers should be read. The $6$--$10$\,mas residuals measure agreement with VIS centroids, while the ${\sim}17$\,mas injection errors measure accuracy against assigned true positions. And the baseline in these comparisons is a classical centroid measured in one band at a time, whereas the head reads the VIS pixels and all ten bands. A classical pipeline could likewise adopt the VIS centroid for every band by cross-matching, and the injection test shows that the head reaches essentially that VIS-limited precision rather than exceeding it. The improvement over single-band centroids therefore measures the value of bringing VIS and the other bands to bear on each source, not a gain over the best classical multi-band position.

Agreement with the training labels could still hide a failure of the estimator itself, and an earlier version of this head without the centroid anchor showed one. Trained on the same features and labels, it learned its own morphology-weighted center convention and displaced a small population of structured galaxies by tens of mas, consistently across bands and across independently trained models. The anchor removes this freedom, since the forward-pass centroid ties every prediction to the label convention within the centroid uncertainty; the affected population shrinks by an order of magnitude, and what remains is confined to faint sources where the bound is wide. Gaia DR3, propagated by proper motion to the effective epoch of the coadds, provides a test independent of the labels: against Gaia, the anchored head places bright stars with the same accuracy as the classical centroids (7.6\,mas median for both, 6.8 versus 6.2\,mas scatter about the frame tie). The head therefore loses nothing on bright stars, whose classical positions are already tied to Gaia through the survey registrations, while retaining its gain of a factor of several for the faint majority. The predicted uncertainty is uniformly conservative across S/N, and a single global rescaling calibrates it as the per-source error bar on the delivered positions.

\subsection{Coherent residual field}

With the per-source scatter under control, the small coherent component can be measured cleanly. Figure~\ref{fig:concordance_field} shows the concordance field, fit to the S/N-weighted offsets with two solvers of deliberately different character: a physics-informed neural network (PINN; \citealt{PINN}) whose auxiliary losses impose curl-free structure, smoothness, and band consistency, and a hierarchical Gaussian process (HGP) that decomposes the field into a component common to all bands, per-instrument group terms, and per-band residuals, with posterior uncertainties; the HGP expands each of these components on a two-scale spatial basis with correlation lengths of $5'$ and $15'$, under prior amplitudes of 4, 2, and 1\,mas for the common, instrument, and band components, while the PINN's spatial resolution is set implicitly by its smoothness penalty rather than by an explicit basis.

The point of using two solvers is that they can fail differently: the PINN constrains the field through physical priors, the HGP through hierarchical smoothness with calibrated posterior uncertainty, and structure that survives both is unlikely to be an artifact of either. On the coarse scales the footprint probes, from the $5'$--$15'$ scales of the HGP basis up to the few-tenths-of-a-degree extent of the field, they recover the same ${\sim}9$--$10$\,mas pattern. Below those scales the comparison is no longer meaningful: refitting the PINN on identical inputs reproduces its own field only to ${\sim}5$\,mas, so finer differences between the solvers are within the fit noise, and we do not interpret structure there.

The $5'$ basis scale is set by the source density. The footprint contains ${\sim}50$--$60$ deduplicated sources per square arcminute (all bands combined), each with tens of mas of centering noise, so the mean offset within a single square arcminute is uncertain by roughly the amplitude of the field itself. Averaging over patches several arcminutes across reduces that uncertainty to $1$--$2$\,mas, at which level the field is clearly measured. The fits therefore cannot reliably resolve structure on finer scales.

The HGP decomposition shows the field is dominated by its common component (${\sim}7$\,mas), shared across all ten bands: most of it is a single cross-survey systematic between the two Gaia-tied solutions rather than per-band noise. Gaia DR3 positions are accurate to ${\sim}0.02$\,mas for bright stars and to ${\sim}1$\,mas even at the $G\approx21$ faint limit \citep{gaia2023}, so an offset of this size cannot originate in the reference frame. It must arise in the step where each survey ties its solution to that frame: the matching of detected stars to Gaia, the centroid errors of those stars in each survey's images, the propagation of Gaia proper motions to the survey epochs, and the residuals of each distortion model. Stated chromatically, the field is nearly achromatic: for a typical Rubin band the common term carries roughly $80\%$ of the field variance, pointing to a geometric origin, a registration offset between the delivered frames, rather than a wavelength-dependent one. The fitted per-band terms sit at only $1.5$--$3$\,mas across $ugrizy$, constraining band dependence on these spatial scales for the selected sources, although they do not bound all source-color-dependent atmospheric effects. The larger NISP per-band residuals track the resampling excess already noted for Figure~\ref{fig:astrom_agreement}, with a possible contribution from the localized persistence-related residuals noted there.

The field is stable to the source selection: refitting with only sources at S/N\,${>}\,10$ changes the per-component amplitudes by less than ${\sim}0.5$\,mas and leaves the spatial pattern correlated at ${>}0.96$ with the fiducial fit, so the field is not an artifact of the faint, high-scatter majority. Refitting the field to the head-corrected offsets gives an amplitude of only ${\sim}1$\,mas: the per-source corrections remove the coherent offset along with the scatter. The fitted field is also consistent with the bright-source offsets of Figure~\ref{fig:astrom_agreement}: subtracting it at the bright-source positions reduces the displacement of their offset distributions from ${\sim}8$ to ${\sim}1$\,mas, with the common component accounting for nearly all of the change, and this holds at positions excluded from the fit provided the excluded gaps are small compared with the $5'$--$15'$ correlation scales. The bright-source offset and the mapped field are therefore the same systematic measured in two ways, and because the fit is dominated by faint sources, bright and faint sources share it.

The concordance field is nonetheless a measurement more than a correction, in two specific senses. For the faint majority, the ${\sim}50$\,mas per-source scatter buries the ${\sim}10$\,mas coherent term, so subtracting the field leaves the pooled median offset unchanged. And the field is coherent but not constant: its local value drifts across the footprint, so it does not extrapolate. Refit with an entire sky patch excluded, it fails to re-center the offset distribution of that patch's bright sources. Its value is as a map of how well two independently calibrated surveys agree on the sky, at the few-mas level on the coarse scales this footprint probes.

\section{Out-of-distribution transfer}\label{sec:ood}

\begin{figure*}[t]
    \centering
    \includegraphics[width=\textwidth]{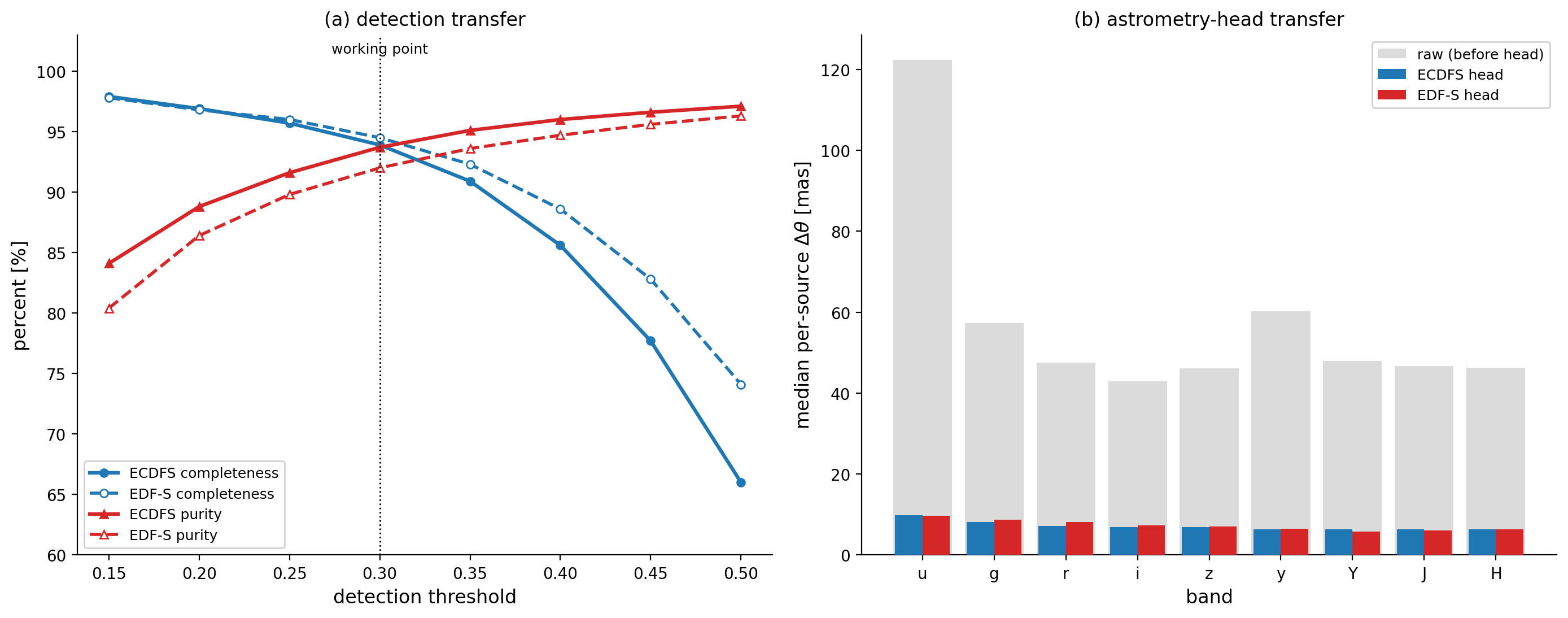}
\caption{Out-of-distribution transfer to EDF-S with no retraining. The frozen foundation and the detection and astrometry heads, trained only on the ECDFS field, are applied to the held-out EDF-S field (72 tiles, a separate \textit{Euclid} Deep Field never seen in training). \emph{(a) Detection:} completeness and catalog-based purity versus detection threshold, in-distribution (ECDFS, solid) and out-of-distribution (EDF-S, dashed). Completeness uses each field's clean VIS-selected \textit{Euclid} Q1 MER sample with VIS\,$<24.5$; purity uses counterparts in the full MER catalog. At the $0.30$ working point EDF-S reaches completeness/purity of $94.5\%/92.0\%$ against ECDFS $93.9\%/93.7\%$: completeness transfers to within $1\%$, and catalog-based purity is lower by ${\sim}2$ points, within the field-to-field variation expected from MER's own incompleteness, which bounds this statistic from below. \emph{(b) Astrometry:} per-band head-corrected median cross-survey offset, ECDFS versus EDF-S; the gray bars mark the raw pre-head offsets. The astrometry head collapses the raw ${\sim}45$--$60$\,mas offsets to the same ${\sim}6$--$10$\,mas floor on the unseen field.}
    \label{fig:ood}
\end{figure*}

The final question is whether any of this transfers. Everything so far (the reconstruction, the detection head, the astrometry head, the concordance field) was measured on ECDFS, the field the foundation was pretrained on. A representation that only works where it was trained is of limited use; the point of a foundation model is that the frozen features carry general structure. We test this directly on EDF-S, the Euclid Deep Field South: a widely separated region of sky, processed identically but never seen during pretraining or head training. Nothing is changed: the same frozen foundation, the same detection and astrometry heads exactly as trained on ECDFS, applied to the 72 EDF-S tiles with no retraining or fine-tuning (Figure~\ref{fig:ood}).

Detection transfers with no loss in completeness. Running the production CenterNet head at the same $0.30$ working point, completeness for clean VIS-selected sources with VIS\,$<24.5$ in EDF-S is $94.5\%$, matching the $93.9\%$ measured in-distribution, and the completeness-versus-threshold curve tracks the ECDFS one across the full range (Figure~\ref{fig:ood}a). Catalog-based purity against the full MER catalog is $92.0\%$, measured within the MER catalog footprint, about $2$ points below the in-distribution $93.7\%$. That gap should not be read as a transfer loss: catalog-based purity is bounded from below by the reference catalog's incompleteness (Section~\ref{sec:detection}), and EDF-S carries its own MER selection function, so a difference of this size is within what the reference itself can account for. The truth-based depth, which does not depend on MER, transfers as well: repeating the star-donor injection protocol of Section~\ref{sec:detection} on the EDF-S tiles covered by its MER catalog yields a $50\%$ point-source depth of VIS\,${\approx}26.8$ ($26.6$ from VIS alone), fully consistent with the in-distribution $26.7$ ($26.4$). Detection therefore carries over to a new field with no measurable loss in completeness or depth, whether scored against the reference catalog or against injected truth.

The astrometry head transfers just as cleanly. Applied to EDF-S, the raw cross-survey offsets are $\sim$45--60\,mas in the Rubin optical bands, comparable to ECDFS, and the head collapses them to per-band medians of ${\sim}7$--$9$\,mas ($g/r/i/z$) and ${\sim}6$\,mas (NISP $Y\!/\!J\!/\!H$), with $u$ highest at ${\sim}10$\,mas, all within ${\sim}1$\,mas of the in-distribution floor (Figure~\ref{fig:ood}b). The transfer also holds against truth rather than labels: sources of known position injected into the EDF-S tiles are recovered to ${\sim}17$\,mas at $\mathrm{S/N}=5$, within ${\sim}1$\,mas of the local VIS anchor floor ($16.1$\,mas), reproducing the in-distribution injection result on sky the head has never seen. The alignment learned on one field applies unchanged to another, which is expected if the head is reading a genuine, instrument-level relationship between the bands rather than a field-specific pattern. The anchor contributes to that robustness by construction: it is computed from the pixels of each new field rather than learned from the training one, so the convention the head is tied to travels with the data.

Two points are worth making about what this does and does not show. It is a same-instrument, same-processing transfer test: EDF-S is a different patch of sky, not different hardware or a different pipeline, so it probes generalization across the sky rather than across surveys. And it uses only currently public Q1 data over a single Rubin tract in each of two deep fields, a small fraction of the available Rubin--\textit{Euclid} overlap. That the frozen stack transfers across that gap with no retraining is what makes scaling the same approach to larger areas, and to the deeper releases to come, a matter of applying it rather than rebuilding it.

\section{Public products}\label{sec:release}

The full pipeline is public at \url{https://github.com/xoubish/JAISP} under a BSD 3-clause license: the foundation model, both heads, the training recipes, the injection and validation scripts, and the notebooks that produce every figure in this paper. The trained weights of the foundation and of the detection and astrometry heads, together with the detection lists and head-corrected per-source positions for the ECDFS and EDF-S fields, are released with the tagged \texttt{v1.0} version of the repository under a CC-BY 4.0 license and archived on Zenodo (doi:\href{https://doi.org/10.5281/zenodo.23598660}{10.5281/zenodo.23598660}). All inputs are public survey products, Rubin DP1 coadds served by the Rubin Science Platform and \textit{Euclid} Q1 MER mosaics served by IRSA \citep{IRSA601}, so the tiles used here can be regenerated with the fetch scripts and tile-center table in the repository. Inference does not require a GPU: the frozen encoder runs in about a minute per tile ($102''$ on a side, Section~\ref{sec:data}) on eight CPU cores and in a fraction of a second on a single GPU with under 5\,GB of memory, and the two heads add negligible cost on top of its cached features, so a field the size of ECDFS takes minutes on a GPU and of order a day on a CPU. The scripts that produced the detection lists and positions in this paper serve as the inference examples.

Users should keep the scope of these products in mind. The foundation was trained on one Rubin tract at DP1 and Q1 depth and has been tested only on a second field with the same instruments and processing; the detection head is VIS-anchored, and the astrometry head delivers positions in the VIS-centroid convention rather than an improved tie to Gaia. The weights are tied to the data releases they were trained on, and as deeper Rubin and \textit{Euclid} releases arrive they should be regarded as a baseline to fine-tune or retrain on the new pixels rather than a final product.

\section{Summary and Discussion}\label{sec:discussion}

This work set out to test whether the joint pixel content of Rubin and \textit{Euclid} can be distilled, without supervision, into one representation precise enough for measurement. The foundation model, trained solely to predict each band of the ECDFS imaging from the rest, clears the three checks posed in Section~\ref{sec:validation}: held-out bands from both instruments are reproduced with high fidelity, per-band brightness reads out linearly from the frozen features, and reconstruction draws on other bands in proportions that track real spectral relationships. On that footing, the two heads deliver their measurements. The detection head reaches the classical VIS baseline and improves on it in two respects: it reduces source confusion, with less fragmentation of extended galaxies in the examples examined, and the full ten-band stack deepens the census by ${\approx}0.3$\,mag over VIS alone, to a $50\%$ point-source depth of VIS\,${\approx}26.7$ ($25.2$ for moderately extended profiles), an increment the native-stem control traces to the fused features rather than the head design. The astrometry head equips every band with VIS-grade positions: the ${\sim}45$\,mas raw disagreement between matched sources shrinks to ${\sim}6$--$10$\,mas, and the error against injected truth at $\mathrm{S/N}=5$ is ${\sim}17$\,mas, nearly saturating the ${\sim}16$\,mas precision of the VIS anchor. For bright stars, whose classical positions are limited by each survey's registration to Gaia, the centroid anchor keeps the head at the same accuracy (7.6\,mas against Gaia for both), so one estimator serves bright and faint sources alike. Separating off the per-source noise exposes the concordance field: a smooth ${\sim}9$--$10$\,mas disagreement between two astrometric calibrations that are each tied to Gaia yet offset from one another, the bulk of it a single ${\sim}7$\,mas term common to all ten bands, a relative offset between the two surveys' delivered frames rather than a band-dependent effect, with per-band terms of only $1.5$--$3$\,mas. The whole stack then repeats these results on a second deep field it has never seen, with nothing retrained. In science terms, the ${\sim}0.3$\,mag deeper ten-band census adds of order $20\%$ more sources at the faint limit and, more importantly, recovers blended neighbors that bias weak-lensing shapes and photometric redshifts; the ${\sim}10$\,mas cross-band positions meet the concordance requirement for joint photometry \citep{Chary2020} and, for the faint majority beyond \textit{Gaia}'s limit, make proper motions and astrometric microlensing measurable across the Rubin--\textit{Euclid} epoch baseline.

All of this is measured on the earliest public data from both observatories, which sets a floor rather than a ceiling. Rubin DP1 is commissioning imaging with a median PSF of roughly $1''$; operational LSST will deliver sharper images and, over the survey, far deeper stacks. Both improvements propagate directly through the pipeline: a sharper PSF tightens the classical centroids that serve as training labels and as raw baselines alike, and deeper optical imaging may give the Rubin bands enough weight to seed detections by themselves, which at DP1 depth they cannot. \textit{Euclid} Q1 will likewise be superseded by the deeper and better-processed DR1, which is expected to deliver the native-sampling NISP mosaics with the variance maps our inputs require (Section~\ref{sec:data}). If the ${\sim}3$\,mas NISP excess of Section~\ref{sec:astrometry} does enter with the resampling, it is exactly the kind of inherited preprocessing residual a joint model could then remove, by taking those mosaics as input and solving the registration within the model. The same reasoning extends earlier in the pipeline: everything here operates on coadded mosaics, and pushing the representation toward single exposures, where more of the instrumental and atmospheric signal survives, is where the joint approach has the most room to contribute.

The relation between the learned stack and the classical pipelines it builds on is complementary rather than competitive. Classical processing supplied every label and every benchmark in this paper (the survey WCS solutions, the \textsc{SEP} detection labels, the Gaussian-centroid targets, the MER cross-checks), and classical understanding dictated the evaluations that made the learned components trustworthy. When joint processing of these surveys was scoped as a classical pipeline effort, the estimate ran to roughly two hundred work-years of development and upward of a billion CPU hours of computing \citep{Chary2020}; that scope covered the entire overlapping footprint of the three surveys, of order thousands of square degrees, and a community science platform, far more than the two heads and single tract demonstrated here, so the numbers do not compare one to one. What the learned stack changes is the cost structure. Pretraining took ${\sim}8$ GPU-hours, the features are extracted once per dataset, and each new capability is a small head trained on the cached features, under a million parameters for both heads presented here; inference is a single forward pass per tile. This division of labor, one expensive representation feeding many cheap heads, is the practical content of the foundation-model paradigm for survey pipelines.

Several of the in-distribution numbers are limited by the references they are measured against rather than by representational capacity. Catalog-based purity describes agreement with the full MER catalog and is bounded by that catalog's incompleteness, which is why detection depth is measured instead by injection, with chance recovery constrained by a near-zero-flux control and unmatched induced detections tracked separately through $P_{\rm inj}$. The astrometric floor is likewise set by the VIS Gaussian-centroid labels the head is trained toward: the injection test places the head within ${\sim}1$\,mas of that anchor, so the head has extracted nearly all the precision its labels contain. This suggests that better labels and anchors may improve measurements with the same architecture: deeper VIS stacks, ePSF-based centroiding, or Gaia-anchored positions provide concrete avenues to test whether that floor can be lowered.

The limitations identified along the way point in a consistent direction. The detection head is VIS-anchored: the other bands add depth in combination with VIS but rarely drive detections on their own, and training labels not derived solely from VIS would be the natural way to relax that. The transfer test of Section~\ref{sec:ood} probes a new field, not new hardware: transfer across instruments, with \textit{Roman} \citep{Spergel2015} as the natural third stream alongside Rubin and \textit{Euclid}, is the harder and more valuable test ahead. The nearest-term step, however, is completing the measurement layer: photometry, redshift, and shape heads, now in development, will work from the same frozen features and treat the per-band morphology and PSF structure the encoder was forced to retain as the measured quantities themselves rather than as aids to localization. Together with the detection and astrometry heads presented here, these would make the joint Rubin--\textit{Euclid} pixels directly usable for the object-level measurements that precision cosmology requires, extracted from all ten bands at once: a single representation, learned once from public imaging, serving every measurement built on it.

\begin{acknowledgments}
This work was supported in part by NASA grants to JPL and Caltech. The analysis makes use of LSST Data Preview 1 data, obtained with the NSF--DOE Vera C. Rubin Observatory during commissioning of the LSST Commissioning Camera; Rubin Observatory is a joint initiative of the U.S.\ National Science Foundation and the U.S.\ Department of Energy Office of Science. It also makes use of data from Quick Data Release 1 of the \textit{Euclid} mission of the European Space Agency (ESA); \textit{Euclid} is an ESA mission with instruments and data provided by the Euclid Consortium. The \textit{Euclid} data were accessed through the NASA/IPAC Infrared Science Archive (IRSA), which is funded by NASA and operated by the California Institute of Technology. Claude (Anthropic) was used to edit and format the manuscript text and to assist with the development and documentation of the public code; all text, analysis, and results were reviewed and verified by the authors, who take full responsibility for the final manuscript.
\end{acknowledgments}
\software{Astropy \citep{astropy2022}, astroquery \citep{astroquery2019}, NumPy \citep{numpy2020}, SciPy \citep{scipy2020}, Matplotlib \citep{matplotlib2007}, pandas \citep{pandas2010}, PyTorch \citep{pytorch}, scikit-learn \citep{sklearn2011}, SEP \citep{Bertin1996,SEP}, LSST Science Pipelines \citep{Bosch2018}, Weights \& Biases}

\bibliographystyle{aasjournalv7}
\bibliography{refs}

\end{document}